\documentclass[preprint,aps,prd,nofootinbib]{revtex4-2}

\usepackage[T1]{fontenc}
\usepackage[utf8]{inputenc}
\usepackage{amsmath,amssymb,bm}
\usepackage{graphicx}
\graphicspath{{./figures/}}
\usepackage{xcolor}
\usepackage[colorlinks=true,linkcolor=blue,citecolor=blue,urlcolor=blue]{hyperref}
\usepackage{float}
\usepackage{placeins}
\usepackage{subcaption}
\usepackage{slashed}
\usepackage{caption}
\begin{document}
		
		\title{The Higgs boson decay $h \to bs$ in the Generational Three-Higgs-Doublet Model}
		
		\author{Hao-Ran Ma$^{1,2,3}$}
		\email{1248186282@qq.com}

		\author{Ti-Bin Hou$^{1,2,3}$}
		\email{3078849770@qq.com}

		\author{Jin-Lei Yang$^{1,2,3}$}
		\email{jlyang@hbu.edu.cn}

		\author{Tai-Fu Feng$^{1,2,3,4}$}
		\email{fengtf@hbu.edu.cn}
		
		\affiliation{$^1$Department of Physics, Hebei University, Baoding 071002, China}
		\affiliation{$^2$Hebei Key Laboratory of High-precision Computation and Application of Quantum Field Theory, Baoding 071002, China}
		\affiliation{$^3$Hebei Research Center of the Basic Discipline for Computational Physics, Baoding 071002, China}
		\affiliation{$^4$Department of Physics, Chongqing University, Chongqing 401331, China}
		
		\begin{abstract}
			The Generational Three-Higgs-Doublet Model (G3HDM) extends the Standard Model (SM) by three Higgs doublets, which couple separately to different generations of quarks and charged leptons, and consequently make significant contributions to the neutral flavor-changing processes. Considering the latest experimental constraints on  $125\mathrm{GeV}$ Higgs signals, electroweak precision observables, neutral-meson mixing, and the rare decay processes $B\to X_s\gamma$ and $B_s^0\to\mu^+\mu^-$, we perform a systematic analysis of the processes $h \to bs$ predicted in the G3HDM. This process is strongly suppressed by the loop factor in the SM, hence any observation of $h \to bs$ in the near future provides a sensitive probe of physics beyond the SM. Finally, the constraints on the parameter space and the experimental testability of G3HDM through observation $h \to bs$ are presented. 
			
		\end{abstract}

		\maketitle
		\clearpage   
	
\section{Introduction}
            Since the experimental discovery of the Higgs boson~\cite{ATLAS:2012qbp,CMS:2012qbp,Evans:2012,Hao:2012,Ibe:2012,Moroi:2012,Ren:2018}, one of the central tasks in particle physics has been to test its physical properties as precisely as possible. So far, within the combined range of experimental uncertainties and theoretical errors, the various measurements of the Higgs boson remain overall consistent with the predictions of the Standard Model (SM). Nevertheless, Higgs flavor-changing interactions are still regarded as one of the important avenues for probing new physics. In the SM, flavor-changing neutral current (FCNC) can only be generated through loop diagrams, and their effects are therefore usually highly suppressed~\cite{Arco:2023}. The latest analyses, combined with current experimental data, indicate that within the SM framework, the branching ratio of the decay process $\mathrm{BR}(h \to bs)$ is only of the order of $10^{-7}$~\cite{BenitezGuzman:2015,Blankenburg:2012,Aranda:2020,Barducci:2017}. However, in many theoretical frameworks beyond the SM, the branching ratio of $h\to bs$ can be significantly enhanced~\cite{Cao:2014,Cao:2017,Gao:2024}. For this reason, any significant deviation observed in such processes could be an indication of new physics contributions beyond the SM. Studies of the loop-induced $hbs$ coupling can be traced back to the 1980s~\cite{Willey:1982mc}. 
            
            The Large Hadron Collider (LHC) experiments currently have very limited direct sensitivity on this process~\cite{Blankenburg:2012,HerreroGarcia:2020}. The dominant decay mode of the Higgs boson is $h \to b\bar b$, and when one of the $b$ jets is mistagged, this process can be misidentified as $h \to bs$. In addition, it is experimentally difficult to distinguish jets originating from $s$ quarks and $d$ quarks. Therefore, the signals of Higgs decays into $bs$ and $bd$ are practically indistinguishable in actual observations. On the other hand, the large Quantum Chromodynamics (QCD) background also poses substantial difficulties for the analysis of processes involving bottom quarks. This makes the measurement of decay processes such as $h \to bj$ (with $j=s,d$) highly challenging at the LHC, and even at the High-Luminosity Large Hadron Collider (HL-LHC). One of the core physics goals of the International Linear Collider (ILC) is to precisely determine the properties of the Higgs boson. For the branching ratios of the Higgs decays $h \to bs$ and $h \to bd$, the ILC is expected, in principle, to reach a sensitivity at the subpercent level~\cite{Barducci:2017}. Although these decay channels are difficult, or even impossible, to probe effectively at the LHC due to the large QCD background, they may be studied more precisely at the ILC thanks to its much cleaner collider environment.

            In the Generational Three-Higgs-Doublet Model (G3HDM), neutral scalars can mediate FCNC interactions at tree level. Therefore, the same flavor structure responsible for enhancing $h\to bs$ also contributes to $\Delta F=2$ observables through neutral-scalar exchange. In particular, it can modify $B_s^0-\bar B_s^0$, $B_d^0-\bar B_d^0$, and $K^0-\bar K^0$ mixing, which are characterized by $\Delta M_{B_s}$, $\Delta M_{B_d}$, $\Delta M_K$, and $\epsilon_K$. Since the experimental measurements of these meson-mixing observables are in good agreement with the SM predictions, they impose stringent constraints on the tree-level FCNC couplings and the neutral-scalar sector of the model~\cite{ParticleDataGroup:2024cfk}. In addition, the extended scalar sector contributes to rare $B$-meson decays with $\Delta F=1$. The decay $\bar B\to X_s\gamma$ is particularly sensitive to charged-Higgs loop contributions, whereas $B_s^0\to\mu^+\mu^-$ may receive additional contributions from neutral scalars. Since the experimental measurements of these processes are also in good agreement with the SM predictions~\cite{Adel:1994,Ali:1995,Greub:1996,Chetyrkin:1997,Misiak:2007,Misiak:2007b,Buras:2012,Misiak:2015,Czakon:2015}, a consistent and comprehensive study of $h\to bs$ in the G3HDM must simultaneously impose the constraints from the $B$ meson rare decays $\bar B\to X_s\gamma$, $B_s^0\to\mu^+\mu^-$ which have been measured precisely~\cite{Lees:2012,Lees:2012b,Saito:2015,BHH1,BHH2,BHH3,ParticleDataGroup:2024cfk}.
             \begin{equation}\label{eq1}
             	\begin{aligned}[b]
             		&\mathrm{Br}(\bar{B} \to X_s \gamma)_{\mathrm{exp}} = (3.49 \pm 0.19) \times 10^{-4}, \\
             		&\mathrm{BR}(B_s \to \mu^+\mu^-)_{\mathrm{exp}} = (3.01 \pm 0.35)\times 10^{-9}.
             	\end{aligned}
             \end{equation}
            All the experimental constraints mentioned above are taken into account in the numerical computations.

            The paper is organized as follows. In Section~\ref{sec:G3HDM}, we briefly introduce the theoretical framework of the G3HDM and present the basic forms of the scalar and fermion mass matrices. In Section~\ref{sec:Pr analysis}, we derive the analytical expression for the branching ratio of $h\to bs$, present the effective Hamiltonians and branching-ratio expressions for $\bar{B}\to X_s\gamma$ and $B_s^0\to\mu^+\mu^-$, and specify the allowed ranges imposed on the electroweak precision observables and the neutral-meson mixing observables $\Delta M_{B_d}$, $\Delta M_{B_s}$, $\Delta M_K$, and $\epsilon_K$. In Section~\ref{sec:Nuanalysis}, we describe the parameter settings adopted in the numerical analysis and present the corresponding numerical results and discussion. Finally, the main conclusions of this work are summarized in Section~\ref{sec:Conclusion}.

            \section{The Generational Three-Higgs-Doublet Model}
            \label{sec:G3HDM}
            
            In this work, we consider a G3HDM, in which the SM Higgs sector is extended by two additional scalar $SU(2)_L$ doublets.
            The model is motivated by the possibility that fermion mass hierarchies (or part thereof) originate not only from hierarchical Yukawa couplings, but also from a hierarchical pattern of electroweak symmetry breaking.
            In particular, the three Higgs doublets may be arranged such that they provide masses predominantly to the first, second, and third generations of SM fermions, respectively, and a hierarchy among their vacuum expectation values can then partially address the SM flavor puzzle.
            
            \subsection{The scalar sector of the G3HDM}
            \label{subsec:G3HDM_scalar}
            
            \paragraph{Field content and scalar potential.}
            The scalar sector contains three Higgs doublets $\Phi_a$ ($a=1,2,3$), transforming under the SM gauge group $SU(3)_c\times SU(2)_L\times U(1)_Y$ as
            \begin{equation}
            	\Phi_a\sim (1,2,\tfrac{1}{2}),
            	\qquad a=1,2,3.
            	\label{eq:Phi_rep}
            \end{equation}
            We assume that the renormalizable Higgs potential respects, to a good approximation, a softly broken $U(1)^3$ symmetry, with each $U(1)$ factor acting on a single Higgs doublet.
            The resulting potential can be written as
            \begin{align}
            	V_{3\mathrm{HDM}}
            	&=
            	m_{11}^2(\Phi_1^\dagger \Phi_1)
            	+ m_{22}^2(\Phi_2^\dagger \Phi_2)
            	+ m_{33}^2(\Phi_3^\dagger \Phi_3)
            	\nonumber\\
            	&\quad
            	-\Big[
            	m_{12}^2(\Phi_1^\dagger \Phi_2)
            	+ m_{23}^2(\Phi_2^\dagger \Phi_3)
            	+ m_{13}^2(\Phi_1^\dagger \Phi_3)
            	+ \mathrm{h.c.}
            	\Big]
            	\nonumber\\
            	&\quad
            	+ \lambda_1(\Phi_1^\dagger \Phi_1)^2
            	+ \lambda_2(\Phi_2^\dagger \Phi_2)^2
            	+ \lambda_3(\Phi_3^\dagger \Phi_3)^2
            	+ \lambda_4(\Phi_1^\dagger \Phi_1)(\Phi_2^\dagger \Phi_2)
            	+ \lambda_5(\Phi_1^\dagger \Phi_1)(\Phi_3^\dagger \Phi_3)
            	\nonumber\\
            	&\quad
            	+ \lambda_6(\Phi_2^\dagger \Phi_2)(\Phi_3^\dagger \Phi_3)
            	+ \lambda_7(\Phi_1^\dagger \Phi_2)(\Phi_2^\dagger \Phi_1)
            	+ \lambda_8(\Phi_1^\dagger \Phi_3)(\Phi_3^\dagger \Phi_1)
            	+ \lambda_9(\Phi_2^\dagger \Phi_3)(\Phi_3^\dagger \Phi_2).
            	\label{eq:V3HDM}
            \end{align}
            Here the diagonal mass parameters $m_{aa}^2$ and all quartic couplings $\lambda_i$ are taken to be real, while the off-diagonal mass parameters $m_{ab}^2$ ($a\neq b$) softly break the approximate $U(1)^3$ symmetry and can in general be complex, potentially inducing CP violation in the Higgs sector.
            
            \paragraph{Electroweak symmetry breaking and VEV parametrization.}
            We assume that electroweak symmetry breaking proceeds as usual,
            $SU(2)_L\times U(1)_Y\to U(1)_{\mathrm{em}}$, with the vacuum aligned such that $U(1)_{\mathrm{em}}$ remains unbroken.
            In a component-field decomposition, one may parametrize
            \begin{equation}
            	\Phi_a=
            	\begin{pmatrix}
            		\varphi_a^+ \\
            		\dfrac{1}{\sqrt{2}}\left(v_a+\phi_a+i a_a\right)
            	\end{pmatrix},
            	\qquad a=1,2,3,
            	\label{eq:Phi_decomp}
            \end{equation}
            where $v_a$ denote the vacuum expectation values (VEVs).
            We work in a phase convention in which the vevs are real and label the fields such that $v_1\ll v_2\ll v_3$.
            The electroweak scale is fixed by
            \begin{equation}
            	v^2\equiv v_1^2+v_2^2+v_3^2\simeq (246~\mathrm{GeV})^2.
            	\label{eq:vsum}
            \end{equation}
            A convenient parametrization trades $(v_1,v_2,v_3)$ for $(v,\beta,\beta')$,
            \begin{equation}
            	v_1=v\cos\beta',
            	\qquad
            	v_2=v\sin\beta'\cos\beta,
            	\qquad
            	v_3=v\sin\beta'\sin\beta,
            	\label{eq:vev_angles}
            \end{equation}
            implying
            \begin{equation}
            	\tan\beta'=\frac{\sqrt{v_2^2+v_3^2}}{v_1},
            	\qquad
            	\tan\beta=\frac{v_3}{v_2}.
            	\label{eq:tanb_def}
            \end{equation}
            
            \paragraph{Tadpole conditions.}
            Working with real vevs generally implies nontrivial relations among the imaginary parts of the soft mass parameters.
            The minimization conditions yield
            \begin{equation}
            	\mathrm{Im}(m_{13}^2)=-\frac{v_2}{v_3}\,\mathrm{Im}(m_{12}^2),
            	\qquad
            	\mathrm{Im}(m_{23}^2)=\frac{v_1}{v_3}\,\mathrm{Im}(m_{12}^2).
            	\label{eq:Im_relations}
            \end{equation}
            The remaining minimization conditions can be used to eliminate the diagonal mass parameters in favor of the vevs
            \begin{align}
            	&m_{11}^2=
            	\mathrm{Re}(m_{12}^2)\frac{v_2}{v_1}
            	+\mathrm{Re}(m_{13}^2)\frac{v_3}{v_1}
            	-\lambda_1 v_1^2
            	-\frac{1}{2}\Big[(\lambda_4+\lambda_7)v_2^2+(\lambda_5+\lambda_8)v_3^2\Big],
            	\label{eq:tad_m11}\\[2mm]
            	&m_{22}^2=
            	\mathrm{Re}(m_{12}^2)\frac{v_1}{v_2}
            	+\mathrm{Re}(m_{23}^2)\frac{v_3}{v_2}
            	-\lambda_2 v_2^2
            	-\frac{1}{2}\Big[(\lambda_4+\lambda_7)v_1^2+(\lambda_6+\lambda_9)v_3^2\Big],
            	\label{eq:tad_m22}\\[2mm]
            	&m_{33}^2=
            	\mathrm{Re}(m_{13}^2)\frac{v_1}{v_3}
            	+\mathrm{Re}(m_{23}^2)\frac{v_2}{v_3}
            	-\lambda_3 v_3^2
            	-\frac{1}{2}\Big[(\lambda_5+\lambda_8)v_1^2+(\lambda_6+\lambda_9)v_2^2\Big].
            	\label{eq:tad_m33}
            \end{align}
            In the following, we adopt the simplifying assumption that the tree-level Higgs potential is CP invariant and set
            \begin{equation}
            	\mathrm{Im}(m_{12}^2)=\mathrm{Im}(m_{13}^2)=\mathrm{Im}(m_{23}^2)=0.
            	\label{eq:CPinv_assumption}
            \end{equation}
            
            \paragraph{Scalar mass matrices and diagonalization.}
            In the CP-conserving limit, the physical spectrum after electroweak symmetry breaking consists of three neutral CP-even scalars $(h,H,H^{\prime})$, two neutral  CP-odd scalars $(A_{min},A_{max})$, and two pairs of charged Higgs bosons $(H_{min}^\pm,H_{max}^\pm)$, in addition to the Goldstone bosons $(G^0,G^\pm)$.
            
            In the basis $(a_1,a_2,a_3)$, the CP-odd mass-squared matrix is independent of quartic couplings and is given by
            \begin{equation}
            	\widehat{m}_a^2=
            	\begin{pmatrix}
            		m_{12}^2\dfrac{v_2}{v_1}+m_{13}^2\dfrac{v_3}{v_1} & -m_{12}^2 & -m_{13}^2 \\
            		-m_{12}^2 & m_{12}^2\dfrac{v_1}{v_2}+m_{23}^2\dfrac{v_3}{v_2} & -m_{23}^2 \\
            		-m_{13}^2 & -m_{23}^2 & m_{13}^2\dfrac{v_1}{v_3}+m_{23}^2\dfrac{v_2}{v_3}
            	\end{pmatrix}.
            	\label{eq:ma2_hat}
            \end{equation}
            The charged and CP-even mass-squared matrices can be expressed compactly as
            \begin{align}
            	&\widehat{m}_\pm^2=
            	\widehat{m}_a^2
            	+\frac{1}{2}
            	\begin{pmatrix}
            		-v_2^2\lambda_7-v_3^2\lambda_8 & v_1v_2\lambda_7 & v_1v_3\lambda_8 \\
            		v_1v_2\lambda_7 & -v_1^2\lambda_7-v_3^2\lambda_9 & v_2v_3\lambda_9 \\
            		v_1v_3\lambda_8 & v_2v_3\lambda_9 & -v_1^2\lambda_8-v_2^2\lambda_9
            	\end{pmatrix},
            	\label{eq:mpm2_hat}\\[2mm]
            	&\widehat{m}_\phi^2=
            	\widehat{m}_a^2
            	+
            	\begin{pmatrix}
            		2v_1^2\lambda_1 & v_1v_2(\lambda_4+\lambda_7) & v_1v_3(\lambda_5+\lambda_8) \\
            		v_1v_2(\lambda_4+\lambda_7) & 2v_2^2\lambda_2 & v_2v_3(\lambda_6+\lambda_9) \\
            		v_1v_3(\lambda_5+\lambda_8) & v_2v_3(\lambda_6+\lambda_9) & 2v_3^2\lambda_3
            	\end{pmatrix}.
            	\label{eq:mphi2_hat}
            \end{align}
            The CP-odd, charged, and CP-even mass-squared matrices can be diagonalized by the unitary mixing matrices $Z^A$, $Z^{\pm}$, and $Z^H$, respectively
            \begin{align}
            	&Z^{A}\,\widehat{m}_a^{2}\,Z^{A\dagger}=
            	\mathrm{diag}\!\left(0,M_{A_{min}}^{2},M_{A_{max}}^{2}\right),
            	\label{eq:diag_A}\\
            	&Z^{\pm}\,\widehat{m}_\pm^{2}\,Z^{\pm\dagger}=
            	\mathrm{diag}\!\left(0,M_{H_{min}^{\pm}}^{2},M_{H_{max}^{\pm}}^{2}\right),
            	\label{eq:diag_Hpm}\\
            	&Z^{H}\,\widehat{m}_\phi^{2}\,Z^{H\dagger}=
            	\mathrm{diag}\!\left(M_{h}^{2},M_{H}^{2},M_{H^{\prime}}^{2}\right).
            	\label{eq:diag_h}
            \end{align}
            where, $Z^A$, $Z^{\pm}$, and $Z^H$ can be written as
            
            \begin{align}
            	Z^A
            	&=
            	\begin{pmatrix}
            		0 & 0 & 1 \\
            		\sin\gamma_A & \cos\gamma_A & 0 \\
            		\cos\gamma_A & -\sin\gamma_A & 0
            	\end{pmatrix}
            	\begin{pmatrix}
            		\sin\beta' & 0 & -\cos\beta' \\
            		0 & 1 & 0 \\
            		\cos\beta' & 0 & \sin\beta'
            	\end{pmatrix}
            	\begin{pmatrix}
            		1 & 0 & 0 \\
            		0 & \sin\beta & -\cos\beta \\
            		0 & \cos\beta & \sin\beta
            	\end{pmatrix},
            	\label{eq:OA-ascending}
            	\\[1ex]
            	Z^{\pm}
            	&=
            	\begin{pmatrix}
            		0 & 0 & 1 \\
            		\sin\gamma_{\pm} & \cos\gamma_{\pm} & 0 \\
            		\cos\gamma_{\pm} & -\sin\gamma_{\pm} & 0
            	\end{pmatrix}
            	\begin{pmatrix}
            		\sin\beta' & 0 & -\cos\beta' \\
            		0 & 1 & 0 \\
            		\cos\beta' & 0 & \sin\beta'
            	\end{pmatrix}
            	\begin{pmatrix}
            		1 & 0 & 0 \\
            		0 & \sin\beta & -\cos\beta \\
            		0 & \cos\beta & \sin\beta
            	\end{pmatrix},
            	\label{eq:Opm-ascending}
            	\\[1ex]
            	Z^H
            	&=
            	\begin{pmatrix}
            		0 & 0 & 1 \\
            		\sin\gamma_H & \cos\gamma_H & 0 \\
            		\cos\gamma_H & -\sin\gamma_H & 0
            	\end{pmatrix}
            	\begin{pmatrix}
            		\cos\alpha' & 0 & \sin\alpha' \\
            		0 & 1 & 0 \\
            		-\sin\alpha' & 0 & \cos\alpha'
            	\end{pmatrix}
            	\begin{pmatrix}
            		1 & 0 & 0 \\
            		0 & \cos\alpha & \sin\alpha \\
            		0 & -\sin\alpha & \cos\alpha
            	\end{pmatrix}.
            	\label{eq:OH-ascending}
            \end{align}

            After applying the $\beta$ and $\beta'$ rotations, one obtains partially diagonalized forms of the respective mass matrices, with the massless Goldstone bosons $G^0$ and $G^\pm$ already appearing as eigenstates. The final rotations by the angles $\gamma_A$ and $\gamma_\pm$ fully diagonalize the mass matrices and define the massive eigenstates $A_{min}$, $A_{max}$ and $H_{min}^\pm$, $H_{max}^{\pm}$. We can use these relations to trade the Lagrangian parameters $m_{12}^2$, $m_{13}^2$, and $m_{23}^2$ for $M_{A_{min}}$, $M_{A_{max}}$, and $\gamma_A$. The charged Higgs masses $M_{H_{min}^{\pm}}$ and $M_{H_{max}^{\pm}}$, as well as the corresponding mixing angle $\gamma_\pm$, can be determined analogously.
            \paragraph{Perturbative unitarity and vacuum stability.}
            
            Imposing the tree-level perturbative unitarity bounds, we adopt the following parameter ranges
            
            \begin{equation}
            	|\lambda_1|,\,|\lambda_2|,\,|\lambda_3| \le \frac{4\pi}{3}, \qquad
            	|\lambda_4 + \lambda_7|,\,|\lambda_5 + \lambda_8|,\,|\lambda_6 + \lambda_9| \le 8\pi.
            	\label{eq:1}
            \end{equation}
            
            Furthermore, vacuum stability requires the scalar potential to be bounded from below, leading to the following conditions
            
            \begin{equation}
            	\begin{aligned}
            		\text{(a)}\quad & \lambda_1>0,\;\lambda_2>0,\;\lambda_3>0,\\[4pt]
            		\text{(b)}\quad & \Lambda_{12}+2\sqrt{\lambda_1\lambda_2}\ge0,\;
            		\Lambda_{13}+2\sqrt{\lambda_1\lambda_3}\ge0,\;
            		\Lambda_{23}+2\sqrt{\lambda_2\lambda_3}\ge0,\\[6pt]
            		\text{(c)}\quad & \sqrt{\lambda_1\lambda_2\lambda_3}
            		+\frac{1}{2}\Big(\Lambda_{12}\sqrt{\lambda_3}
            		+\Lambda_{13}\sqrt{\lambda_2}
            		+\Lambda_{23}\sqrt{\lambda_1}\Big) \\
            		& +\frac{1}{2}\sqrt{\Big(\Lambda_{12}+2\sqrt{\lambda_1\lambda_2}\Big)
            			\Big(\Lambda_{13}+2\sqrt{\lambda_1\lambda_3}\Big)
            			\Big(\Lambda_{23}+2\sqrt{\lambda_2\lambda_3}\Big)} \ge 0,
            	\end{aligned}
            	\label{eq:2}
            \end{equation}
            
            where
            \begin{equation}
            	\Lambda_{12} = \lambda_4 + \min(0,\lambda_7),\quad
            	\Lambda_{13} = \lambda_5 + \min(0,\lambda_8),\quad
            	\Lambda_{23} = \lambda_6 + \min(0,\lambda_9).
            	\label{eq:3}
            \end{equation}
            
            \subsection{The fermion sector in the G3HDM}
            \label{subsec:G3HDM_fermion}
            
            The Yukawa interactions of the three Higgs doublets $\Phi_a$ ($a=1,2,3$) with the SM fermions are described by the most general renormalizable Yukawa Lagrangian
            \begin{align}
            	&-\mathcal{L}^{\mathrm{Yuk}}_{3\mathrm{HDM}}=
            	\sum_{a=1}^{3}\sum_{i,j=1}^{3}
            	\Big(
            	\lambda^{ua}_{ij}\,\bar q_{Li}\,\widetilde{\Phi}_a\,u_{Rj}
            	+\lambda^{da}_{ij}\,\bar q_{Li}\,\Phi_a\,d_{Rj}
            	+\lambda^{\ell a}_{ij}\,\bar \ell_{Li}\,\Phi_a\,e_{Rj}
            	\Big)
            	+\mathrm{h.c.},
            	\label{eq:Yukawa_Lagrangian}
            \end{align}
            where $\widetilde{\Phi}_a\equiv i\sigma_2\Phi_a^{\ast}$ and $i,j$ are flavor indices.
            Neutrino masses and mixing are neglected in the present setup.
            
            To realize a \emph{generational} structure, we adopt Yukawa textures of the form~\cite{Altmannshofer:2016zrn,Altmannshofer:2017uvs,Altmannshofer:2018bch}
            \begin{subequations}\label{eq:Yukawa_textures}
            	\begin{align}
            		&\lambda_{u1} \sim \frac{\sqrt{2}}{v_1}
            		\begin{pmatrix}
            			m_u & m_u & m_u\\
            			m_u & m_u & m_u\\
            			m_u & m_u & m_u
            		\end{pmatrix},
            		\qquad
            		\lambda_{u2} \sim \frac{\sqrt{2}}{v_2}
            		\begin{pmatrix}
            			0 & 0 & 0\\
            			0 & m_c & m_c\\
            			0 & m_c & m_c
            		\end{pmatrix},
            		\qquad
            		\lambda_{u3} \sim \frac{\sqrt{2}}{v_3}
            		\begin{pmatrix}
            			0 & 0 & 0\\
            			0 & 0 & 0\\
            			0 & 0 & m_t
            		\end{pmatrix},
            		\label{eq:Yukawa_textures_u}\\[2mm]
            		&\lambda_{d1} \sim \frac{\sqrt{2}}{v_1}
            		\begin{pmatrix}
            			m_d & m_s\,\lambda & m_b\,\lambda^3\\
            			m_d & m_d & m_d\\
            			m_d & m_d & m_d
            		\end{pmatrix},
            		\qquad
            		\lambda_{d2} \sim \frac{\sqrt{2}}{v_2}
            		\begin{pmatrix}
            			0 & 0 & 0\\
            			0 & m_s & m_b\,\lambda^2\\
            			0 & m_s & m_s
            		\end{pmatrix},
            		\qquad
            		\lambda_{d3} \sim \frac{\sqrt{2}}{v_3}
            		\begin{pmatrix}
            			0 & 0 & 0\\
            			0 & 0 & 0\\
            			0 & 0 & m_b
            		\end{pmatrix},
            		\label{eq:Yukawa_textures_d}\\[2mm]
            		&\lambda_{\ell 1} \sim \frac{\sqrt{2}}{v_1}
            		\begin{pmatrix}
            			m_e & m_e & m_e\\
            			m_e & m_e & m_e\\
            			m_e & m_e & m_e
            		\end{pmatrix},
            		\qquad
            		\lambda_{\ell 2} \sim \frac{\sqrt{2}}{v_2}
            		\begin{pmatrix}
            			0 & 0 & 0\\
            			0 & m_\mu & m_\mu\\
            			0 & m_\mu & m_\mu
            		\end{pmatrix},
            		\qquad
            		\lambda_{\ell 3} \sim \frac{\sqrt{2}}{v_3}
            		\begin{pmatrix}
            			0 & 0 & 0\\
            			0 & 0 & 0\\
            			0 & 0 & m_\tau
            		\end{pmatrix}.
            		\label{eq:Yukawa_textures_l}
            	\end{align}
            \end{subequations}
            Here ``$\sim$'' indicates the order of magnitude of the entries; nonzero entries within a given matrix may differ by complex $\mathcal{O}(1)$ factors.
            Moreover, $\lambda\simeq |V_{us}|$ denotes the Wolfenstein parameter.
            All Yukawa matrices are assumed to be rank-1; consequently, each Higgs doublet couples only to a single linear combination of the three fermion generations, which constitutes the core idea of the G3HDM.
            
            Assuming that the CKM matrix originates from the diagonalization of the down-quark mass matrix, we define in the fermion mass-eigenstate basis the following mass parameters
            \begin{align}
            	&m^{f1}_{ff'} = \frac{v_1}{\sqrt{2}}\,\langle f_L|\lambda_{f1}|f'_R\rangle,
            	\qquad
            	m^{f2}_{ff'} = \frac{v_2}{\sqrt{2}}\,\langle f_L|\lambda_{f2}|f'_R\rangle,
            	\qquad
            	m^{f3}_{ff'} = \frac{v_3}{\sqrt{2}}\,\langle f_L|\lambda_{f3}|f'_R\rangle,
            	\label{eq:mf_def}
            \end{align}
            which satisfy
            \begin{equation}
            	m^{f3}_{ff'}+m^{f2}_{ff'}+m^{f1}_{ff'}=m_f\,\delta_{ff'}.
            	\label{eq:mf_sumrule}
            \end{equation}
            with $m_f$ denoting the physical fermion masses.
            
            Expanding to leading order in the ratios of first-to-second and second-to-third generation masses, one obtains~\cite{Altmannshofer:2025pjj}
            \begin{align}
            	&\frac{m^{u1}_{qq'}}{m_u} \simeq
            	\begin{pmatrix}
            		1 & O^{u}_{uc} & O^{u}_{ut}\\
            		O^{u}_{cu} & O^{u}_{cu}O^{u}_{uc} & O^{u}_{cu}O^{u}_{ut}\\
            		O^{u}_{tu} & O^{u}_{tu}O^{u}_{uc} & O^{u}_{tu}O^{u}_{ut}
            	\end{pmatrix},
            	\qquad
            	\frac{m^{u2}_{qq'}}{m_c} \simeq
            	\begin{pmatrix}
            		\dfrac{m_u^2}{m_c^2}O^{u}_{uc}O^{u}_{cu} & -\dfrac{m_u}{m_c}O^{u}_{uc} & -\dfrac{m_u}{m_c}O^{u}_{uc}O^{u}_{ct}\\
            		-\dfrac{m_u}{m_c}O^{u}_{cu} & 1 & O^{u}_{ct}\\
            		-\dfrac{m_u}{m_c}O^{u}_{tc}O^{u}_{cu} & O^{u}_{tc} & O^{u}_{tc}O^{u}_{ct}
            	\end{pmatrix},
            	\label{eq:mu12}\\[1mm]
            	&\frac{m^{u3}_{qq'}}{m_t} \simeq
            	\begin{pmatrix}
            		\dfrac{m_u^2}{m_t^2}(O^{u}_{ut}-O^{u}_{uc}O^{u}_{ct})(O^{u}_{tu}-O^{u}_{tc}O^{u}_{cu}) & \dfrac{m_um_c}{m_t^2}(O^{u}_{ut}-O^{u}_{uc}O^{u}_{ct})O^{u}_{tc} & -\dfrac{m_u}{m_t}(O^{u}_{ut}-O^{u}_{uc}O^{u}_{ct})\\
            		\dfrac{m_um_c}{m_t^2}O^{u}_{ct}(O^{u}_{tu}-O^{u}_{tc}O^{u}_{cu}) & \dfrac{m_c^2}{m_t^2}O^{u}_{ct}O^{u}_{tc} & -\dfrac{m_c}{m_t}O^{u}_{ct}\\
            		-\dfrac{m_u}{m_t}(O^{u}_{tu}-O^{u}_{tc}O^{u}_{cu}) & -\dfrac{m_c}{m_t}O^{u}_{tc} & 1
            	\end{pmatrix},
            	\label{eq:mu3}\\[1mm]
            	&\frac{m^{d1}_{qq'}}{m_d} \simeq
            	\begin{pmatrix}
            		1 & \dfrac{m_s}{m_d}V_{ud}^\ast V_{us} & \dfrac{m_b}{m_d}V_{ud}^\ast V_{ub}\\
            		O^{d}_{sd} & O^{d}_{sd}\dfrac{m_s}{m_d}V_{ud}^\ast V_{us} & O^{d}_{sd}\dfrac{m_b}{m_d}V_{ud}^\ast V_{ub}\\
            		O^{d}_{bd} & O^{d}_{bd}\dfrac{m_s}{m_d}V_{ud}^\ast V_{us} & O^{d}_{bd}\dfrac{m_b}{m_d}V_{ud}^\ast V_{ub}
            	\end{pmatrix},
            	\qquad
            	\frac{m^{d2}_{qq'}}{m_s} \simeq
            	\begin{pmatrix}
            		-\dfrac{m_d}{m_s}V_{cd}^\ast V_{cs}\,O^{d}_{sd} & V_{cd}^\ast V_{cs} & \dfrac{m_b}{m_s}V_{cd}^\ast V_{cb}\\
            		-\dfrac{m_d}{m_s}O^{d}_{sd} & 1 & \dfrac{m_b}{m_s}V_{cs}^\ast V_{cb}\\
            		-\dfrac{m_d}{m_s}O^{d}_{bs}O^{d}_{sd} & O^{d}_{bs} & O^{d}_{bs}\dfrac{m_b}{m_s}V_{cs}^\ast V_{cb}
            	\end{pmatrix},
            	\label{eq:md12}\\[1mm]
            	&\frac{m^{d3}_{qq'}}{m_b} \simeq
            	\begin{pmatrix}
            		-\dfrac{m_d}{m_b}V_{td}^\ast V_{tb}(O^{d}_{bd}-O^{d}_{bs}O^{d}_{sd}) & -\dfrac{m_s}{m_b}V_{td}^\ast V_{tb}O^{d}_{bs} & V_{td}^\ast V_{tb}\\
            		-\dfrac{m_d}{m_b}V_{ts}^\ast V_{tb}(O^{d}_{bd}-O^{d}_{bs}O^{d}_{sd}) & -\dfrac{m_s}{m_b}V_{ts}^\ast V_{tb}O^{d}_{bs} & V_{ts}^\ast V_{tb}\\
            		-\dfrac{m_d}{m_b}(O^{d}_{bd}-O^{d}_{bs}O^{d}_{sd}) & -\dfrac{m_s}{m_b}O^{d}_{bs} & 1
            	\end{pmatrix}.
            	\label{eq:md3}
            \end{align}
            The lepton mass parameters are completely analogous to those in the up quark sector. In the above expressions, $O^{q}_{ij}$ are free, in general complex, $\mathcal{O}(1)$ parameters, encoding additional sources of flavor and CPV beyond the SM.
            
            \section{Process analysis}
            \label{sec:Pr analysis}
            
            \subsection{Amplitude and branching ratio of $h \to bs$}
            \label{subsec:hbs zfu}
            
\begin{figure}[H]
            	\centering
            	
            	\begin{subfigure}{0.24\textwidth}
            		\centering
            		\includegraphics[width=\linewidth]{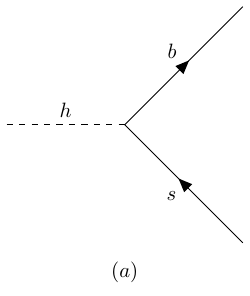}
            	\end{subfigure}
            	
            	\vspace{0.4cm}
            	\begin{subfigure}{0.24\textwidth}
            		\centering
            		\includegraphics[width=\linewidth]{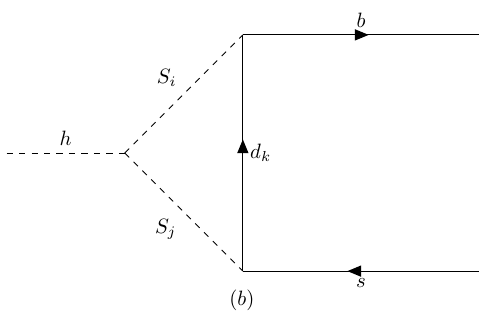}
            	\end{subfigure}
            	\hfill
            	\begin{subfigure}{0.24\textwidth}
            		\centering
            		\includegraphics[width=\linewidth]{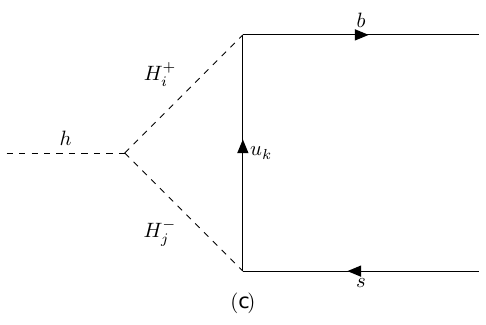}
            	\end{subfigure}
            	\hfill
            	\begin{subfigure}{0.24\textwidth}
            		\centering
            		\includegraphics[width=\linewidth]{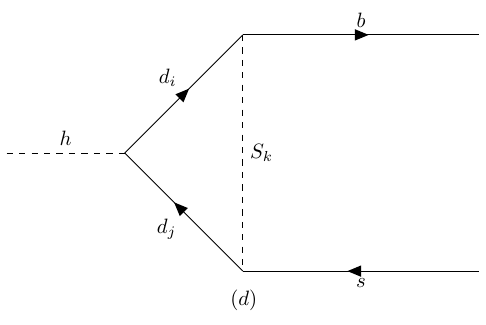}
            	\end{subfigure}
            	\hfill
            	\begin{subfigure}{0.24\textwidth}
            		\centering
            		\includegraphics[width=\linewidth]{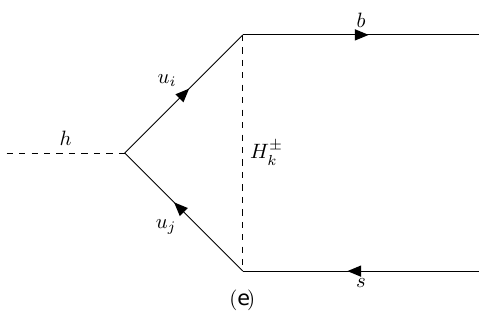}
            	\end{subfigure}
            	
            	\vspace{0.4cm}
            	
            	\begin{subfigure}{0.24\textwidth}
            		\centering
            		\includegraphics[width=\linewidth]{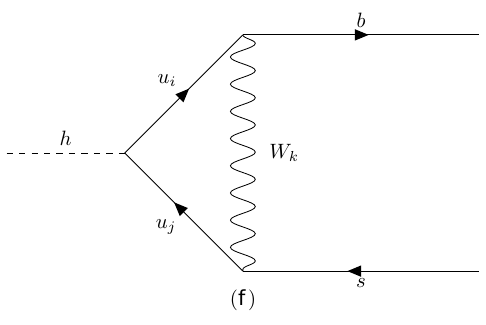}
            	\end{subfigure}
            	\hfill
            	\begin{subfigure}{0.24\textwidth}
            		\centering
            		\includegraphics[width=\linewidth]{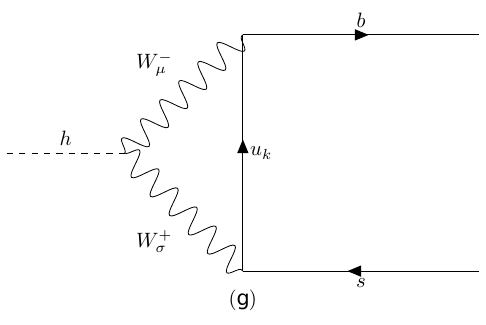}
            	\end{subfigure}
            	\hfill
            	\begin{subfigure}{0.24\textwidth}
            		\centering
            		\includegraphics[width=\linewidth]{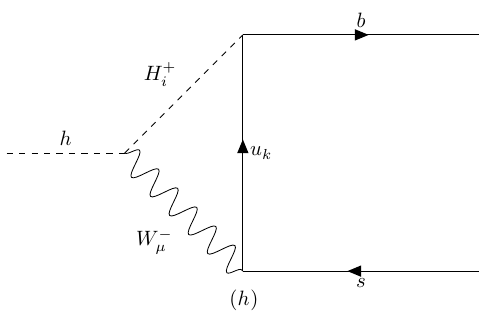}
            	\end{subfigure}
            	\hfill
            	\begin{subfigure}{0.24\textwidth}
            		\centering
            		\includegraphics[width=\linewidth]{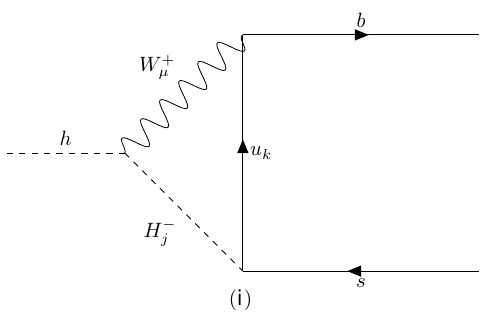}
            	\end{subfigure}
            	
            	\vspace{0.4cm}
            	
            	\begin{subfigure}{0.27\textwidth}
            		\centering
            		\includegraphics[width=\linewidth]{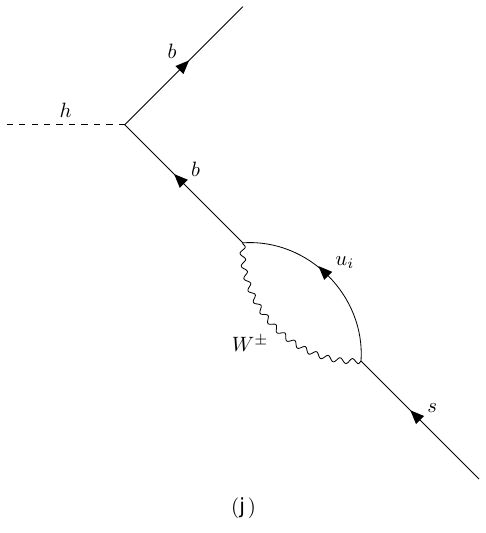}
            	\end{subfigure}
            	\hfill
            	\begin{subfigure}{0.27\textwidth}
            		\centering
            		\includegraphics[width=\linewidth]{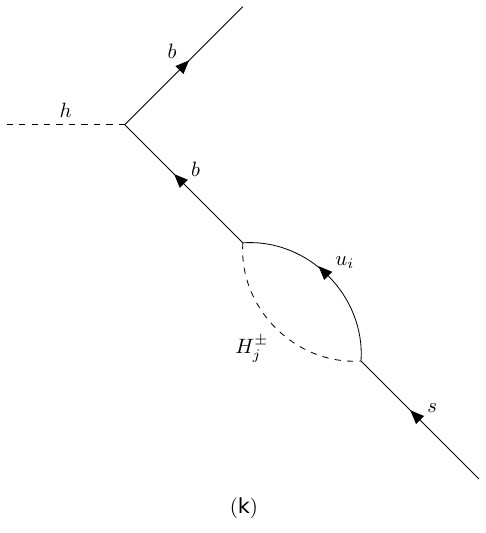}
            	\end{subfigure}
            	\hfill
            	\begin{subfigure}{0.27\textwidth}
            		\centering
            		\includegraphics[width=\linewidth]{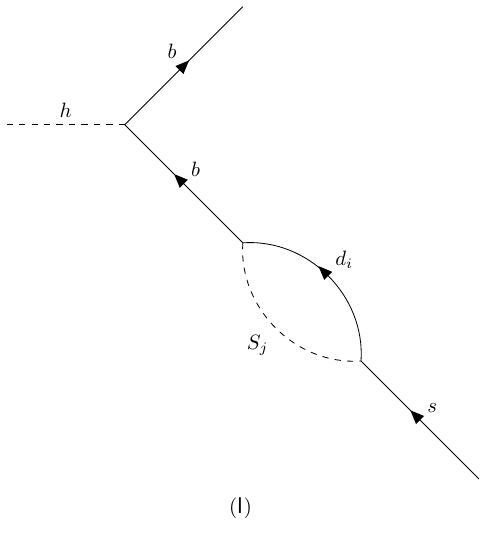}
            	\end{subfigure}
            	
            	\vspace{0.4cm}
            	
            	\begin{subfigure}{0.27\textwidth}
            		\centering
            		\includegraphics[width=\linewidth]{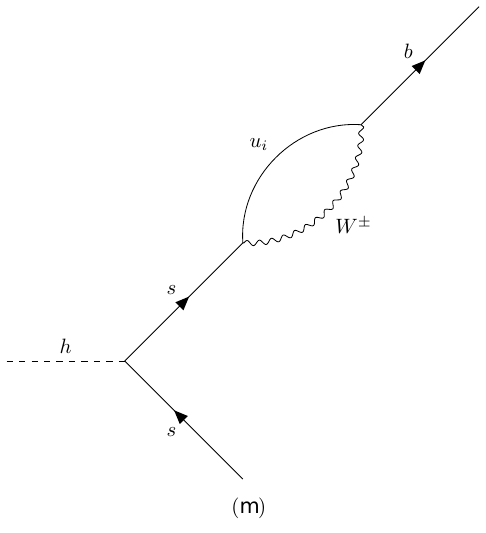}
            	\end{subfigure}
            	\hfill
            	\begin{subfigure}{0.27\textwidth}
            		\centering
            		\includegraphics[width=\linewidth]{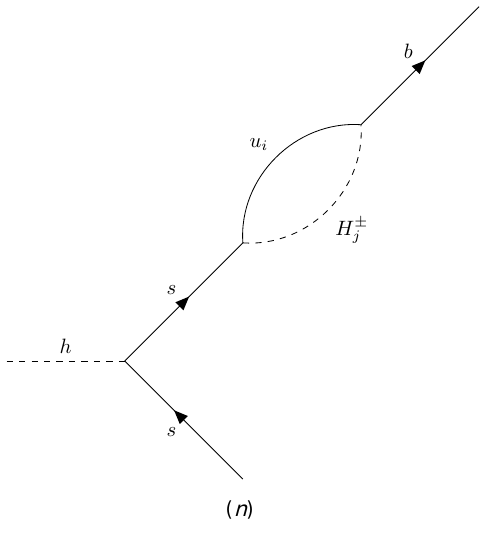}
            	\end{subfigure}
            	\hfill
            	\begin{subfigure}{0.27\textwidth}
            		\centering
            		\includegraphics[width=\linewidth]{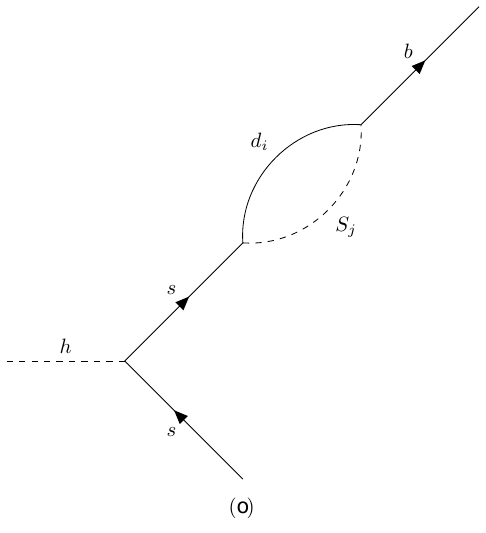}
            	\end{subfigure}
            	
            	\caption{Feynman diagrams for the $h \to \bar{b}s$ process in the G3HDM.}
            	\label{fig:allfigures}
            \end{figure}

            In order to explain how the calculation of the feynman diagrams in Fig.~\ref{fig:allfigures} has been performed, we will take the calculation of Fig.~\ref{fig:allfigures}(b) below for example. The corresponding amplitude can be written as
            
            \begin{equation}\label{eq:M1}
            	\mathcal{M}_{(b)}=
            	\int \frac{d^4k}{(2\pi)^4}\,
            	\frac{
            		\bar u_b(p_b)\,
            		\Bigl(A_L P_L + A_R P_R\Bigr)
            		\bigl(\slashed{k}-\slashed{p_b}+m_{d_k}\bigr)
            		\Bigl(B_L P_L + B_R P_R\Bigr)\,
            		v_s(p_s)\,C_{h{S_i}{S_j}}
            	}{
            		\bigl((k-p_b-p_s)^2-m_{S_j}^2\bigr)\bigl(k^2-m_{S_i}^2\bigr)\bigl((k-p_b)^2-m_{d_k}^2\bigr)
            	}.
            \end{equation}
            where $\bar u_b(p_b)$ and $v_s(p_s)$ denote the Dirac spinors of the bottom and strange quarks, respectively. The quantities $A_{L,R}$, $B_{L,R}$, and $C_{hS_iS_j}$ denote the constant coupling coefficients associated with the interaction vertices $S_i\bar{b}d_k$, $S_j\bar{d}_k s$, and $hS_iS_j$, respectively. The subscripts $L$ and $R$ refer to the left-handed and right-handed chiral components. These vertex factors can be obtained with SARAH~\cite{Staub:2008,Staub:2010,Staub:2011,Staub:2013,Staub:2014}. Using the on shell conditions and the Dirac equations for the external quarks, the amplitude can be simplified as 
            
            \begin{equation}\label{eq:M11}
            	\mathcal{M}_{(b)}=\frac{1}{16\pi^2}\,\bar u_b(p_b)\left(\kappa_{hL}P_L+\kappa_{hR}P_R\right)v_s(p_s).
            \end{equation}
            where 
            
           \begin{equation}\label{eq:kappa1}
           	\begin{aligned}[b]
           		\kappa_{hL}
           		&=
           		i\,C_{h{S_i}{S_j}}\Bigl[
           		A_L B_L\, m_{d_k}\,
           		C_0\!\left(
           		m_b^2, m_s^2, m_h^2,
           		m_{S_i}^2, m_{d_k}^2, m_{S_j}^2
           		\right)
           		\\
           		&\quad
           		+ A_L B_R\, m_s\,
           		C_1\!\left(
           		m_h^2, m_s^2, m_b^2,
           		m_{S_i}^2, m_{S_j}^2, m_{d_k}^2
           		\right)
           		\\
           		&\quad
           		+ A_R B_L\, m_b\,
           		C_1\!\left(
           		m_b^2, m_h^2, m_s^2,
           		m_{d_k}^2, m_{S_i}^2, m_{S_j}^2
           		\right)
           		\Bigr], \\
           		\kappa_{hR}
           		&=
           		\kappa_{hL}(L \leftrightarrow R).
           	\end{aligned}
           \end{equation} 
           Here $C_0$ and $C_1$ are Passarino--Veltman scalar functions~\cite{Denner:1993}. The other diagrams corresponding to $h \to bs$ can be calculated similarly and contribute to the operators $P_L$ and $P_R$.
           
           In the G3HDM, we assume that the only CP-violating effect comes from the CKM matrix, which is known to be very small. Therefore, we compute only the decay width of the process $h \to \bar{b}s$ and multiply it by 2 to obtain the complete prediction for $\Gamma(h \to bs)$. Here, $\Gamma(h \to bs)$ should be understood as the sum $\Gamma(h \to b\bar{s})+\Gamma(h \to \bar{b}s)$. The decay width $\Gamma$ of $h \to bs$ is obtained by substituting $|\mathcal{M}|^2$ into the following formula~\cite{Arco:2023}:
           
           \begin{equation}\label{eq:M^2}
           	\Gamma(h \to bs)=2N_c\,\frac{\lambda^{1/2}\!\left(m_h^2,m_b^2,m_s^2\right)}{16\pi m_h^3}\,|\mathcal{M}|^2.
           \end{equation}
           here, $N_c = 3$ is a color factor, and $\lambda(x,y,z) = (x-y-z)^2 - 4y^2z^2$. The branching ratio we obtained is
           
           \begin{equation}\label{eq:Brhbs}
           	\mathrm{Br}(h \to bs)=\frac{\Gamma(h \to bs)}{\Gamma(h)},
           \end{equation}
           here, $\Gamma(h)=4.1\times10^{-3}\,\mathrm{GeV}$~\cite{LHCHiggsXS:2016}.
           
           \subsection{Theoretical calculation on $\bar{B} \to X_s \gamma$ and $B_s^0 \to \mu^+ \mu^-$}
           \label{subsec:BrBmumu and Brbsr}
           
           The effective Hamiltonian for the transition $b \to s$ at hadronic scale can be written as
           
           \begin{equation}\label{eq:Heff}
           	H_{\mathrm{eff}}
           	=
           	-\frac{4G_F}{\sqrt{2}}\,V_{ts}^*V_{tb}
           	\left[
           	C_1 \mathcal{O}_1^{c}
           	+
           	C_2 \mathcal{O}_2^{c}
           	+
           	\sum_{i=3}^{6} \mathcal{O}_i
           	+
           	\sum_{i=7}^{10}\left(C_i \mathcal{O}_i + C_i' \mathcal{O}_i'\right)
           	+
           	\sum_{i=S,P}\left(C_i \mathcal{O}_i + C_i' \mathcal{O}_i'\right)
           	\right].
           \end{equation}
           where $\mathcal{O}_i,(i=1,2,\ldots,10,S,P)$ and $\mathcal{O}_i',(i=7,8,\ldots,10,S,P)$ are defined as~\cite{Grigjanis:1993,Buchalla:1996,Altmannshofer:2009,Lin:2009,Yang:2010,Goertz:2011} 
           
          \begin{equation}\label{eq:Oeff}
          	\begin{aligned}[b]
          		\mathcal{O}_1^u &= (\bar{s}_L \gamma_\mu T^a u_L)(\bar{u}_L \gamma^\mu T^a b_L), 
          		&\qquad
          		\mathcal{O}_2^u &= (\bar{s}_L \gamma_\mu u_L)(\bar{u}_L \gamma^\mu b_L), \\
          		\mathcal{O}_3 &= (\bar{s}_L \gamma_\mu b_L)\sum_q (\bar{q}\gamma^\mu q), 
          		&\qquad
          		\mathcal{O}_4 &= (\bar{s}_L \gamma_\mu T^a b_L)\sum_q (\bar{q}\gamma^\mu T^a q), \\
          		\mathcal{O}_5 &= (\bar{s}_L \gamma_\mu \gamma_\nu \gamma_\rho b_L)\sum_q (\bar{q}\gamma^\mu \gamma^\nu \gamma^\rho q), 
          		&\qquad
          		\mathcal{O}_6 &= (\bar{s}_L \gamma_\mu \gamma_\nu \gamma_\rho T^a b_L)\sum_q (\bar{q}\gamma^\mu \gamma^\nu \gamma^\rho T^a q), \\
          		\mathcal{O}_7 &= \frac{e}{16\pi^2} m_b (\bar{s}_L \sigma_{\mu\nu} b_R)F^{\mu\nu}, 
          		&\qquad
          		\mathcal{O}_7' &= \frac{e}{16\pi^2} m_b (\bar{s}_R \sigma_{\mu\nu} b_L)F^{\mu\nu}, \\
          		\mathcal{O}_8 &= \frac{g_s}{16\pi^2} m_b (\bar{s}_L \sigma_{\mu\nu} T^a b_R)G^{a,\mu\nu}, 
          		&\qquad
          		\mathcal{O}_8' &= \frac{g_s}{16\pi^2} m_b (\bar{s}_R \sigma_{\mu\nu} T^a b_L)G^{a,\mu\nu}, \\
          		\mathcal{O}_9 &= \frac{e^2}{g_s^2} (\bar{s}_L \gamma_\mu b_L)\,\bar{l}\gamma^\mu l, 
          		&\qquad
          		\mathcal{O}_9' &= \frac{e^2}{g_s^2} (\bar{s}_R \gamma_\mu b_R)\,\bar{l}\gamma^\mu l, \\
          		\mathcal{O}_{10} &= \frac{e^2}{g_s^2} (\bar{s}_L \gamma_\mu b_L)\,\bar{l}\gamma^\mu \gamma_5 l, 
          		&\qquad
          		\mathcal{O}_{10}' &= \frac{e^2}{g_s^2} (\bar{s}_R \gamma_\mu b_R)\,\bar{l}\gamma^\mu \gamma_5 l, \\
          		\mathcal{O}_S &= \frac{e^2}{16\pi^2} m_b (\bar{s}_L b_R)\,\bar{l}l, 
          		&\qquad
          		\mathcal{O}_S' &= \frac{e^2}{16\pi^2} m_b (\bar{s}_R b_L)\,\bar{l}l, \\
          		\mathcal{O}_P &= \frac{e^2}{16\pi^2} m_b (\bar{s}_L b_R)\,\bar{l}\gamma_5 l, 
          		&\qquad
          		\mathcal{O}_P' &= \frac{e^2}{16\pi^2} m_b (\bar{s}_R b_L)\,\bar{l}\gamma_5 l,
          	\end{aligned}
          \end{equation}
           where $g_s$ denotes the strong coupling, $F_{\mu\nu}$ are the electromagnetic field strength tensor, $G_{\mu\nu}$ are the gluon field strength tensors, and $T^a$ $(a=1,\ldots,8)$ are the SU(3) generators.
           
\begin{figure}[htbp]
           	\centering
           	\begin{minipage}{0.48\textwidth}
           		\centering
           		\includegraphics[width=\linewidth]{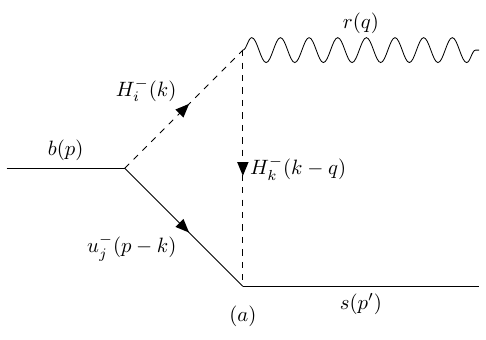}
           	\end{minipage}
           	\hfill
           	\begin{minipage}{0.48\textwidth}
           		\centering
           		\includegraphics[width=\linewidth]{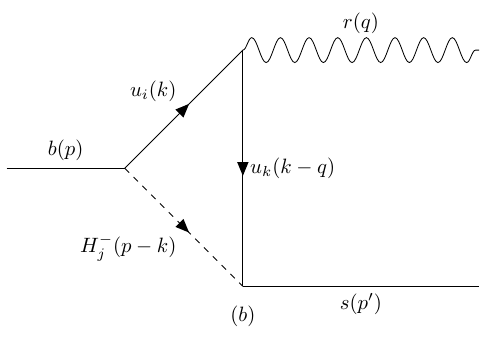}
           	\end{minipage}
           	\caption{\raggedright The one-loop Feynman diagrams contributing to $\bar{B} \to X_s \gamma$ from charged Higgs in the G3HDM.}
           	\label{fig:bsr}
           \end{figure}

           \paragraph{Rare decay $\bar{B}\to X_s\gamma$.}
           Compared with the SM, the main one-loop Feynman diagrams contributing to the process $\bar{B}\to X_s\gamma$ in the G3HDM are shown in Fig.~\ref{fig:bsr}. Then the branching ratio of $\bar{B}\to X_s\gamma$ in the G3HDM can be written as 
           
           \begin{equation}\label{eq:Brbsr}
           	Br(\bar{B}\to X_s\gamma)=R\left(\left|C_{7\gamma}(\mu_b)\right|^2+N(E_\gamma)\right),
           \end{equation}
           where the overall factor $R = 2.47 \times 10^{-3}$, and the nonperturbative contribution $N(E_\gamma) = (3.6 \pm 0.6)\times 10^{-3}$~\cite{Buras:2011}. $C_{7\gamma}(\mu_b)$ is defined by
           
           \begin{equation}\label{eq:C7γ}
           	C_{7\gamma}(\mu_b)=C_{7\gamma,SM}(\mu_b)+C_{7,NP}(\mu_b)\, ,
           \end{equation}
           where we choose the hadron scale $\mu_b = 2.5,\mathrm{GeV}$ and use the SM contribution at NNLO level $C_{7\gamma,\mathrm{SM}}(\mu_b) = -0.3689$~\cite{Gambino:2001,Czakon:2007}. The Wilson coefficients for new physics at the bottom quark scale can be written as~\cite{Buras:1994,Gao:2012}
           
           \begin{equation}\label{eq:C7np}
           	C_{7,NP}(\mu_b)\approx 0.5696\,C_{7,NP}(\mu_{EW})
           	+0.1107\,C_{8,NP}(\mu_{EW}) \, .
           \end{equation}
           where
           
           \begin{equation}\label{eq:C7npew}
           	\begin{aligned}[b]
           		C_{7,NP}(\mu_{EW}) &= C_{7,NP}^{(a)}(\mu_{EW}) + C_{7,NP}^{(b)}(\mu_{EW}) \\
           		&\quad + C_{7,NP}^{\prime (a)}(\mu_{EW}) + C_{7,NP}^{\prime (b)}(\mu_{EW}) \, ,\\
           		C_{8,NP}(\mu_{EW}) &= C_{8g,NP}(\mu_{EW}) + C_{8g,NP}^{\prime}(\mu_{EW}) \, .
           	\end{aligned}
           \end{equation}
           The coefficients $C_{7,\mathrm{NP}}^{(1,2)}(\mu_{EW})$ are Wilson coefficients of the process $b \to s\gamma$ and can be calculated from the diagrams in Fig.~\ref{fig:bsr}(a), (b), respectively, the results read

          	\begin{align}
          		C_{7,NP}^{(a)}(\mu_{EW})
          		&=
          		\sum_{H_i^-,\,u_j}
          		\frac{s_W^2}{2e^2V_{ts}^{*}V_{tb}}
          		\Bigg\{
          		\frac{1}{2}
          		C_{H_i^- \bar{s}u_j}^{R}
          		C_{H_i^- b\bar{u}_j}^{L}
          		\left[
          		-I_3(x_{u_j},x_{H_i^-})+I_4(x_{u_j},x_{H_i^-})
          		\right]
          		\nonumber\\
          		&\qquad\qquad
          		+\frac{m_{u_j}}{m_b}
          		C_{H_i^- \bar{s}u_j}^{L}
          		C_{H_i^- b\bar{u}_j}^{L}
          		\left[
          		-I_1(x_{u_j},x_{H_i^-})+I_3(x_{u_j},x_{H_i^-})
          		\right]
          		\Bigg\},
          		\nonumber\\[1ex]
          		C_{7,NP}^{(b)}(\mu_{EW})
          		&=
          		\sum_{H_j^-,\,u_i}
          		\frac{s_W^2}{3e^2V_{ts}^{*}V_{tb}}
          		\Bigg\{
          		\frac{1}{2}
          		C_{H_j^- \bar{s}u_i}^{R}
          		C_{H_j^- b\bar{u}_i}^{L}
          		\left[
          		-I_1(x_{u_i},x_{H_j^-})
          		+2I_3(x_{u_i},x_{H_j^-})
          		-I_4(x_{u_i},x_{H_j^-})
          		\right]
          		\nonumber\\
          		&\qquad\qquad
          		+\frac{m_{u_i}}{m_b}
          		C_{H_j^- \bar{s}u_i}^{L}
          		C_{H_j^- b\bar{u}_i}^{L}
          		\left[
          		I_1(x_{u_i},x_{H_j^-})
          		-I_2(x_{u_i},x_{H_j^-})
          		-I_3(x_{u_i},x_{H_j^-})
          		\right]
          		\Bigg\},
          		\nonumber\\[1ex]
          		C_{7,NP}^{\prime (\triangle)}(\mu_{EW})
          		&=
          		C_{7,NP}^{(\triangle)}(\mu_{EW})(L\leftrightarrow R),
          		\qquad (\triangle=a,b).
          	\end{align}
           where $x_i = \dfrac{m_i^2}{m_W^2}$, $C_{abc}^{L,R}$ denotes the scalar parts of the interaction vertex about $abc$, with $a$, $b$, $c$ denoting the interactional particles, and the loop integral functions $I_{1,\ldots,4}$ can be found in Appendix~\ref{app:loopfuctions}. In addition, $C_{8g,\mathrm{NP}}(\mu_{EW})$ and $C_{8g,\mathrm{NP}}^{\prime}(\mu_{EW})$ at electroweak scale are
           \begin{equation}
           	\begin{aligned}[b]
           		C_{8g,NP}(\mu_{EW}) &= \left[ C_{7,NP}^{(2)}(\mu_{EW}) + C_{7,NP}^{(3)}(\mu_{EW}) \right] / Q_u, \\
           		C_{8g,NP}^{\prime}(\mu_{EW}) &= C_{8g,NP}(\mu_{EW})(L \leftrightarrow R),
           	\end{aligned}
           \end{equation}
           where $Q_u=\dfrac{2}{3}$.
           
           \paragraph{Rare decay $B_s^0 \to \mu^+\mu^-$.}
           
\begin{figure}[htp]
           	\centering
           	
           	\begin{subfigure}[t]{0.23\textwidth}
           		\centering
           		\includegraphics[width=\linewidth,pagebox=cropbox]{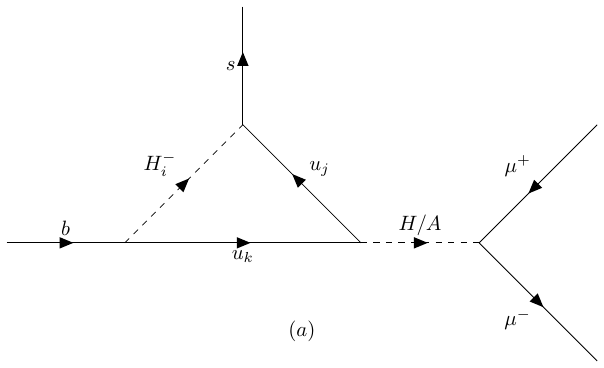}
           	\end{subfigure}
           	\hfill
           	\begin{subfigure}[t]{0.23\textwidth}
           		\centering
           		\includegraphics[width=\linewidth,pagebox=cropbox]{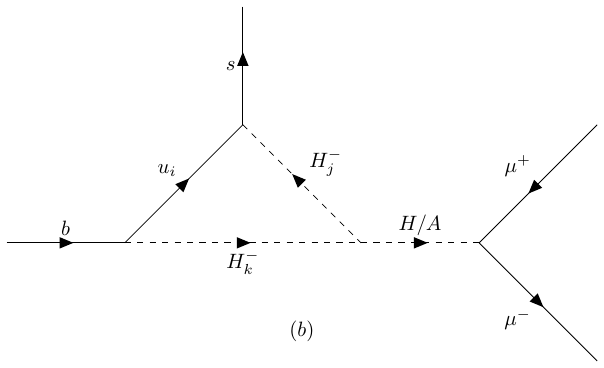}
           	\end{subfigure}
           	\hfill
           	\begin{subfigure}[t]{0.23\textwidth}
           		\centering
           		\includegraphics[width=\linewidth,pagebox=cropbox]{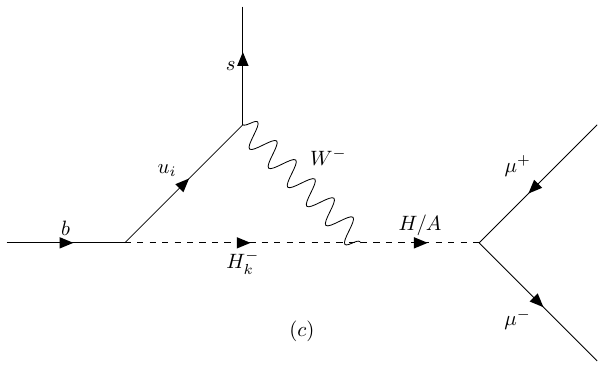}
           	\end{subfigure}
           	\hfill
           	\begin{subfigure}[t]{0.23\textwidth}
           		\centering
           		\includegraphics[width=\linewidth,pagebox=cropbox]{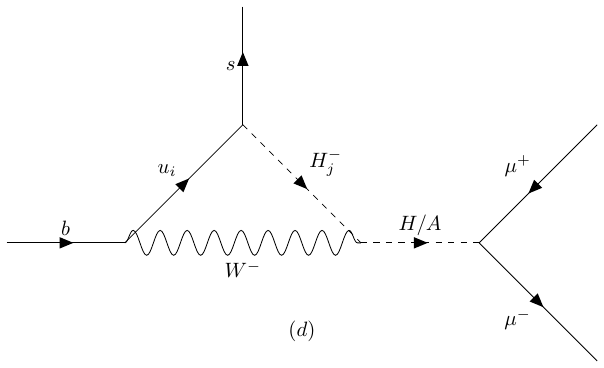}
           	\end{subfigure}
           	
           	\vspace{0.25cm}
           	
           	\begin{subfigure}[t]{0.23\textwidth}
           		\centering
           		\includegraphics[width=\linewidth,pagebox=cropbox]{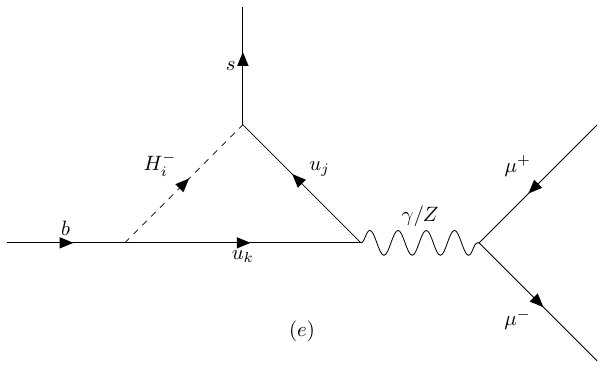}
           	\end{subfigure}
           	\hfill
           	\begin{subfigure}[t]{0.23\textwidth}
           		\centering
           		\includegraphics[width=\linewidth,pagebox=cropbox]{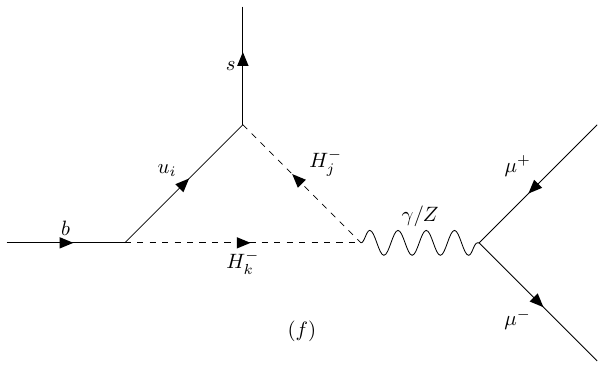}
           	\end{subfigure}
           	\hfill
           	\begin{subfigure}[t]{0.23\textwidth}
           		\centering
           		\includegraphics[width=\linewidth,pagebox=cropbox]{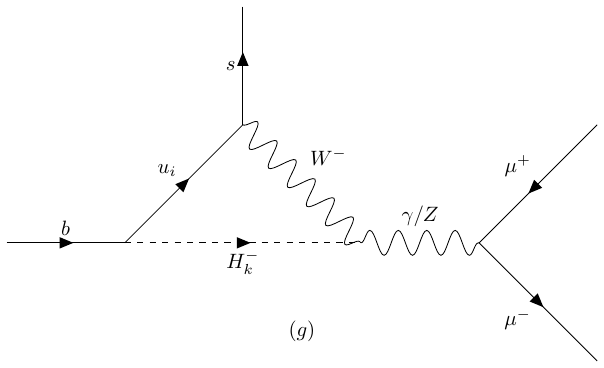}
           	\end{subfigure}
           	\hfill
           	\begin{subfigure}[t]{0.23\textwidth}
           		\centering
           		\includegraphics[width=\linewidth,pagebox=cropbox]{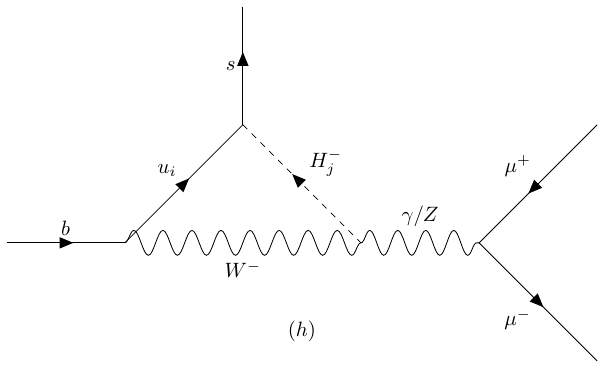}
           	\end{subfigure}
           	
           	\caption{The Feynman diagrams contributing to the decay $B_s^0 \to \mu^+\mu^-$ in the G3HDM.}
           	\label{fig:eightfigs}
           \end{figure}

           The main Feynman diagrams contributing to $B_s^0 \to \mu^+\mu^-$ are plotted in Fig.~\ref{fig:eightfigs}. At the electroweak energy scale $\mu_{EW}$, the corresponding Wilson coefficients can be written as
           
           \begin{align}\label{eq:010}
           	&C_{S,NP}(\mu_{EW})
           	=
           	\frac{\sqrt{2}s_W c_W}{4m_b e^3 V_{ts}^{*}V_{tb}}
           	\Big[
           	C_{S,NP}^{(a)}(\mu_{EW})
           	+
           	C_{S,NP}^{(b)}(\mu_{EW})
           	+
           	C_{S,NP}^{(c)}(\mu_{EW})
           	+
           	C_{S,NP}^{(d)}(\mu_{EW})
           	+
           	C_{S,NP}^{(f)}(\mu_{EW})
           	\Big],
           	\nonumber\\
           	&C_{S,NP}^{\prime}(\mu_{EW})
           	=
           	C_{S,NP}(\mu_{EW})(L\leftrightarrow R),
           	\nonumber\\[1ex]
           	&C_{P,NP}(\mu_{EW})
           	=
           	\frac{\sqrt{2}s_W c_W}{4m_b e^3 V_{ts}^{*}V_{tb}}
           	\Big[
           	C_{P,NP}^{(a)}(\mu_{EW})
           	+
           	C_{P,NP}^{(b)}(\mu_{EW})
           	+
           	C_{P,NP}^{(c)}(\mu_{EW})
           	+
           	C_{P,NP}^{(d)}(\mu_{EW})
           	+
           	C_{P,NP}^{(f)}(\mu_{EW})
           	\Big],
           	\nonumber\\
           	&C_{P,NP}^{\prime}(\mu_{EW})
           	=
           	-\,C_{P,NP}(\mu_{EW})(L\leftrightarrow R),
           	\nonumber\\[1ex]
           	&C_{9,NP}(\mu_{EW})
           	=
           	\frac{\sqrt{2}s_W c_W g_s^2}{64\pi^2 e^3 V_{ts}^{*}V_{tb}}
           	\Big[
           	C_{9,NP}^{(e)}(\mu_{EW})
           	+
           	C_{9,NP}^{(f)}(\mu_{EW})
           	+
           	C_{9,NP}^{(g)}(\mu_{EW})
           	+
           	C_{9,NP}^{(h)}(\mu_{EW})
           	\Big],
           	\nonumber\\
           	&C_{9,NP}^{\prime}(\mu_{EW})
           	=
           	C_{9,NP}(\mu_{EW})(L\leftrightarrow R),
           	\nonumber\\[1ex]
           	&C_{10,NP}(\mu_{EW})
           	=
           	\frac{\sqrt{2}s_W c_W g_s^2}{64\pi^2 e^3 V_{ts}^{*}V_{tb}}
           	\Big[
           	C_{10,NP}^{(e)}(\mu_{EW})
           	+
           	C_{10,NP}^{(f)}(\mu_{EW})
           	+
           	C_{10,NP}^{(g)}(\mu_{EW})
           	+
           	C_{10,NP}^{(h)}(\mu_{EW})
           	\Big],
           	\nonumber\\
           	&C_{10,NP}^{\prime}(\mu_{EW})
           	=
           	-\,C_{10,NP}(\mu_{EW})(L\leftrightarrow R).
           \end{align}
           
           \begin{table}[htbp]
           	\centering
           	\caption{\raggedright At hadronic scale $\mu = m_b$, SM Wilson coefficients to next-to-next-to-logarithmic accuracy}
           	\label{tab:SMWilson}
           	\begin{tabular*}{\textwidth}{@{\extracolsep{\fill}}cccc@{}}
           		\hline\hline
           		$C_7^{eff,SM}$ & $C_8^{eff,SM}$ & $C_9^{eff,SM}$ & $C_{10}^{eff,SM}$ \\
           		\hline
           		$-0.304$ & $-0.167$ & $4.211$ & $-4.103$ \\
           		\hline\hline
           	\end{tabular*}
           \end{table}
           
           The superscripts $(a,\ldots,h)$ correspond respectively to the contributions in Fig.~\ref{fig:eightfigs}$(a,\ldots,h)$. The specific expressions for these Wilson coefficients are detailed in Ref.~\cite{Yang:2018}. The Wilson coefficients at hadronic energy scale from the SM to next-to-next-to-logarithmic accuracy are shown in Table~\ref{tab:SMWilson}. In addition, the Wilson coefficients in Eq.~(\ref{eq:010}) should be evolved down to hadronic scale $\mu \sim m_b$ by the renormalization group equations (Table~\ref{tab:SMWilson}):
           
           \begin{align}\label{eq:C1}
           	\vec{C}_{NP}(\mu) &= \hat{U}(\mu,\mu_0)\,\vec{C}_{NP}(\mu_0), \nonumber\\
           	\vec{C}_{NP}^{\prime}(\mu) &= U^{\prime}(\mu,\mu_0)\,\vec{C}_{NP}^{\prime}(\mu_0).
           \end{align}
           with
           \begin{align}\label{eq:C2}
           	\vec{C}_{NP}^{\,T}
           	&=
           	\left(
           	C_{1,NP},\ \cdots,\ C_{6,NP},\ C_{7,NP}^{eff},\ C_{8,NP}^{NP},
           	\right.
           	\nonumber\\
           	&\quad \left.
           	C_{9,NP}^{eff}-Y(q^2),\ C_{10,NP}^{eff}
           	\right),
           	\nonumber\\
           	\vec{C}_{\mathrm{NP}}^{\prime\,T}
           	&=
           	\left(
           	C_{7,NP}^{\prime,eff},\ C_{8,NP}^{\prime,eff},\ C_{9,NP}^{\prime,eff},\ C_{10,NP}^{\prime,eff}
           	\right).
           \end{align}
           Correspondingly, the evolving matrices are approached as
           \begin{align}
           	\hat{U}(\mu,\mu_0) &\simeq 1-\left[\frac{1}{2\beta_0}\ln\frac{\alpha_s(\mu)}{\alpha_s(\mu_0)}\right]\hat{\gamma}^{(0)T}, \nonumber\\
           	\hat{U}^{\prime}(\mu,\mu_0) &\simeq 1-\left[\frac{1}{2\beta_0}\ln\frac{\alpha_s(\mu)}{\alpha_s(\mu_0)}\right]\hat{\gamma}^{\prime(0)T}.
           \end{align}
           where the anomalous dimension matrices can be read from Ref.~\cite{Gambino:2003} as 
           \begin{equation}
           	\hat{\gamma}^{\mathrm{eff}(0)}=
           	\left(
           	\begin{array}{cccccccccc}
           		-4 & \dfrac{8}{3} & 0 & -\dfrac{2}{9} & 0 & 0 & -\dfrac{208}{243} & \dfrac{173}{162} & -\dfrac{2272}{729} & 0 \\[1ex]
           		12 & 0 & 0 & \dfrac{4}{3} & 0 & 0 & \dfrac{416}{81} & \dfrac{70}{27} & \dfrac{1952}{243} & 0 \\[1ex]
           		0 & 0 & 0 & -\dfrac{52}{3} & 0 & 2 & -\dfrac{176}{81} & \dfrac{14}{27} & -\dfrac{6752}{243} & 0 \\[1ex]
           		0 & 0 & -\dfrac{40}{9} & -\dfrac{100}{9} & \dfrac{4}{9} & \dfrac{5}{6} & -\dfrac{152}{243} & -\dfrac{587}{162} & -\dfrac{2192}{729} & 0 \\[1ex]
           		0 & 0 & 0 & -\dfrac{256}{3} & 0 & 20 & -\dfrac{6272}{81} & \dfrac{6596}{27} & -\dfrac{84032}{243} & 0 \\[1ex]
           		0 & 0 & -\dfrac{256}{9} & \dfrac{56}{9} & \dfrac{40}{9} & -\dfrac{2}{3} & \dfrac{4624}{243} & \dfrac{4772}{81} & -\dfrac{37856}{729} & 0 \\[1ex]
           		0 & 0 & 0 & 0 & 0 & 0 & \dfrac{32}{3} & 0 & 0 & 0 \\[1ex]
           		0 & 0 & 0 & 0 & 0 & 0 & -\dfrac{32}{9} & \dfrac{28}{3} & 0 & 0 \\[1ex]
           		0 & 0 & 0 & 0 & 0 & 0 & 0 & 0 & 0 & 0 \\[1ex]
           		0 & 0 & 0 & 0 & 0 & 0 & 0 & 0 & 0 & 0
           	\end{array}
           	\right).
           \end{equation}
           
           \begin{equation}
           	\hat{\gamma}^{\prime\,\mathrm{eff}(0)}=
           	\left(
           	\begin{array}{cccc}
           		\dfrac{32}{3} & 0 & 0 & 0 \\[1ex]
           		-\dfrac{32}{9} & \dfrac{28}{3} & 0 & 0 \\[1ex]
           		0 & 0 & 0 & 0 \\[1ex]
           		0 & 0 & 0 & 0
           	\end{array}
           	\right).
           \end{equation}
           
           Then, the squared amplitude can be written as
           \begin{equation}
           	|\mathcal{M}_s|^2
           	=
           	16G_F^2\,|V_{tb}V_{ts}^{*}|^2\,M_{B_s^0}^{2}
           	\left[
           	|F_S^s|^2 + |F_P^s + 2m_{\mu}F_A^s|^2
           	\right],
           \end{equation}
           and
           \begin{align}
           	F_S^s &= \frac{\alpha_{EW}(\mu_b)}{8\pi}\,
           	\frac{m_b M_{B_s^0}^2}{m_b+m_s}\,
           	f_{B_s^0}\,(C_S-C_S'), \nonumber\\
           	F_P^s &= \frac{\alpha_{EW}(\mu_b)}{8\pi}\,
           	\frac{m_b M_{B_s^0}^2}{m_b+m_s}\,
           	f_{B_s^0}\,(C_P-C_P'), \nonumber\\
           	F_A^s &= \frac{\alpha_{EW}(\mu_b)}{8\pi}\,
           	f_{B_s^0}\,
           	\left[
           	C_{10}^{\mathrm{eff}}(\mu_b)-C_{10}^{\prime\,\mathrm{eff}}(\mu_b)
           	\right].
           \end{align}
           where $f_{B_s^0}=(227\pm 8),\mathrm{MeV}$ denotes the decay constant, $M_{B_s^0}=5.367,\mathrm{GeV}$ denotes the mass of the neutral meson $B_s^0$. The branching ratio of $B_s^0 \to \mu^+\mu^-$ can be written as
           \begin{equation}
           	\mathrm{Br}(B_s^0 \to \mu^+\mu^-)
           	=
           	\frac{\tau_{B_s^0}}{16\pi}\,
           	\frac{|\mathcal{M}_s|^2}{M_{B_s^0}}\,
           	\sqrt{1-\frac{4m_\mu^2}{M_{B_s^0}^2}} \, .
           \end{equation}
           with $\tau_{B_s^0}=1.466(31) \mathrm{ps}$ denoting the lifetime of the meson.
           
           \subsection{Meson mixing constraints}
           \label{subsec: Meson mixing constraints}
           
            Neutral meson--antimeson mixing is highly sensitive to flavor changing interactions beyond the SM and therefore provides important constraints on possible new physics contributions. In the G3HDM, after rotating the fermion fields to the mass basis, the extended scalar sector and Yukawa structure can induce flavor changing neutral-Higgs interactions. These interactions contribute to $\Delta F=2$ transitions and can therefore affect the neutral-meson mixing observables $\Delta M_{B_d}$, $\Delta M_{B_s}$, $\Delta M_K$, as well as the indirect CP-violating parameter $\epsilon_K$.
           
           In our numerical analysis, we impose the neutral-meson mixing constraints in terms of the ratios of the full theoretical predictions to the corresponding SM reference values. Specifically, a parameter point is retained only if these ratios fall within the following allowed intervals
           \begin{equation}
           	\begin{aligned}
           		&0.8 <
           		\frac{\Delta M_{B_d}}
           		{\Delta M_{B_d}^{\rm SM}}
           		< 1.3,
           		\qquad
           		0.8 <
           		\frac{\Delta M_{B_s}}
           		{\Delta M_{B_s}^{\rm SM}}
           		< 1.3,
           		\\
           		&0 <
           		\frac{\Delta M_K}
           		{\Delta M_K^{\rm SM}}
           		< 2.0,
           		\qquad
           		0.8 <
           		\frac{\epsilon_K}
           		{\epsilon_K^{\rm SM}}
           		< 1.3 .
           	\end{aligned}
           	\label{eq:meson-mixing-constraints}
           \end{equation}
           The above intervals are used as conservative numerical requirements. In particular, a wider range is allowed for $\Delta M_K$ to account for the sizable theoretical uncertainty associated with long-distance contributions.
           
           \subsection{The $125~\mathrm{GeV}$ Higgs boson decays}
           
           The observed $125~\mathrm{GeV}$ Higgs boson provides an important probe of extended scalar sectors. In the G3HDM, this particle is identified with the lightest neutral Higgs mass eigenstate, whose couplings to gauge bosons and fermions may deviate from the corresponding SM predictions. Therefore, precise measurements of the Higgs signal strengths at the LHC can impose important constraints on the parameter space of the G3HDM. The signal strength of the $125~\mathrm{GeV}$ Higgs boson is defined as~\cite{ParticleDataGroup:2024cfk} 
           \begin{align}
           	&\mu_{\gamma\gamma}(h)
           	=
           	\frac{
           		\sigma(gg \to h^{\mathrm{NP}})\,
           		\mathrm{BR}(h^{\mathrm{NP}} \to \gamma\gamma)
           	}{
           		\sigma(gg \to h^{\mathrm{SM}})\,
           		\mathrm{BR}(h^{\mathrm{SM}} \to \gamma\gamma)
           	}
           	= 1.10 \pm 0.06 ,
           	\label{eq:mu_h2_gammagamma}
           	\\
           	&\mu_{WW^*}(h)
           	=
           	\frac{
           		\sigma(gg \to h^{\mathrm{NP}})\,
           		\mathrm{BR}(h^{\mathrm{NP}} \to WW^*)
           	}{
           		\sigma(gg \to h^{\mathrm{SM}})\,
           		\mathrm{BR}(h^{\mathrm{SM}} \to WW^*)
           	}
           	= 1.00 \pm 0.08 ,
           	\label{eq:mu_h2_WW}
           	\\
           	&\mu_{ZZ^*}(h)
           	=
           	\frac{
           		\sigma(gg \to h^{\mathrm{NP}})\,
           		\mathrm{BR}(h^{\mathrm{NP}} \to ZZ^*)
           	}{
           		\sigma(gg \to h^{\mathrm{SM}})\,
           		\mathrm{BR}(h^{\mathrm{SM}} \to ZZ^*)
           	}
           	= 1.02 \pm 0.08 ,
           	\label{eq:mu_h2_ZZ}
           	\\
           	&\mu_{b\bar b}(h)
           	\simeq
           	\frac{
           		\Gamma(h^{\mathrm{NP}} \to VV^*)\,
           		\mathrm{BR}(h^{\mathrm{NP}} \to b\bar b)
           	}{
           		\Gamma(h^{\mathrm{SM}} \to VV^*)\,
           		\mathrm{BR}(h^{\mathrm{SM}} \to b\bar b)
           	}
           	= 0.99 \pm 0.12 ,
           	\label{eq:mu_h2_bb}
           	\\
           	&\mu_{\tau^+\tau^-}(h)
           	\simeq
           	\frac{
           		\Gamma(h^{\mathrm{NP}} \to VV^*)\,
           		\mathrm{BR}(h^{\mathrm{NP}} \to \tau^+\tau^-)
           	}{
           		\Gamma(h^{\mathrm{SM}} \to VV^*)\,
           		\mathrm{BR}(h^{\mathrm{SM}} \to \tau^+\tau^-)
           	}
           	= 0.91 \pm 0.09 .
           	\label{eq:mu_h2_tautau}
           \end{align}
           
           For reference, the corresponding SM predictions are~\cite{ParticleDataGroup:2024cfk} 
           \begin{align}
           	&\Gamma_{\mathrm{tot},125}^{\mathrm{SM}}
           	\simeq 0.0041\,\mathrm{GeV},
           	\label{eq:Gamma_tot_SM_125}
           	\\
           	&\mathrm{BR}(h^{\mathrm{SM}}\to\gamma\gamma)
           	\simeq 0.00227,
           	\label{eq:BR_SM_h2_gammagamma}
           	\\
           	&\mathrm{BR}(h^{\mathrm{SM}}\to WW^*)
           	\simeq 0.214,
           	\label{eq:BR_SM_h2_WW}
           	\\
           	&\mathrm{BR}(h^{\mathrm{SM}}\to ZZ^*)
           	\simeq 0.0262,
           	\label{eq:BR_SM_h2_ZZ}
           	\\
           	&\mathrm{BR}(h^{\mathrm{SM}}\to b\bar b)
           	\simeq 0.582,
           	\label{eq:BR_SM_h2_bb}
           	\\
           	&\mathrm{BR}(h^{\mathrm{SM}}\to\tau^+\tau^-)
           	\simeq 0.0627.
           	\label{eq:BR_SM_h2_tautau}
           \end{align}
           
           In the present NP framework, the total decay width is approximated by 
           \begin{align}
           	&\Gamma_{\mathrm{tot}}^{\mathrm{NP}}(h)
           	\simeq
           	\Gamma(h \to b\bar b)
           	+ \Gamma(h \to c\bar c)
           	+ \Gamma(h \to \tau^+\tau^-)
           	\nonumber
           	\\
           	&\quad
           	+ \sum_{V=W,Z}\Gamma(h \to VV^*)
           	+ \Gamma(h \to gg).
           	\qquad
           	\label{eq:Gamma_tot_NP_hi}
           \end{align}
           where subleading channels such as $\gamma\gamma$, $Z\gamma$, and $\mu^+\mu^-$, as well as possible exotic decay modes, are neglected.
           
           The partial decay widths of the Higgs bosons into fermions and gauge bosons in the G3HDM can be expressed as~\cite{Djouadi:2008,Ellis:1976,Shifman:1979,Bergstrom:1985}
           \begin{align}
           	&\Gamma^{\mathrm{NP}}(h \to gg)
           	=
           	\frac{G_F \alpha_s^2 m_{h}^3}{64\sqrt{2}\pi^3}
           	\left[
           	\left|
           	\sum_{q}
           	g^{S}_{h q\bar q}
           	A_{1/2}(x_q)
           	\right|^2
           	+
           	\left|
           	\sum_{q}
           	g^{A}_{h q\bar q}
           	A_{2}(x_q)
           	\right|^2
           	\right],
           	\label{eq:Gamma_hi_gg}
           	\\[2mm]
           	&\Gamma^{\mathrm{NP}}(h \to \gamma\gamma)
           	=
           	\frac{G_F \alpha^2 m_{h}^3}{128\sqrt{2}\pi^3}
           	\left[
           	\left|
           	\sum_{f=u,d,\ell}
           	N_c^f e_f^2
           	g^{S}_{h f\bar f}
           	A_{1/2}(x_f)
           	+
           	g^{S}_{h WW}
           	A_1(x_W)
           	\right.\right.
           	\nonumber
           	\\
           	&\left.\left.\hspace{3.0cm}
           	+
           	\sum_{a=1}^{2}
           	g^{S}_{h H_a^+H_a^-}
           	\frac{m_Z^2}{m_{H_a^\pm}^2}
           	A_0(x_{H_a^\pm})
           	\right|^2
           	+
           	\left|
           	\sum_{f=u,d,\ell}
           	N_c^f e_f^2
           	g^{A}_{h f\bar f}
           	A_2(x_f)
           	\right|^2
           	\right],
           	\label{eq:Gamma_hi_gammagamma}
           	\\[2mm]
           	&\Gamma^{\mathrm{NP}}(h \to VV^*)
           	=
           	\left|
           	g^{S}_{hVV}
           	\right|^2
           	\Gamma^{\mathrm{SM}}(h \to VV^*),
           	\qquad V=W,Z ,
           	\label{eq:Gamma_hi_VV}
           	\\[2mm]
           	&\Gamma^{\mathrm{NP}}(h \to f\bar f)
           	=
           	N_c^f
           	\frac{G_F m_f^2 m_{h}}{4\sqrt{2}\pi}
           	\left[
           	\left|
           	g^{S}_{h f\bar f}
           	\right|^2
           	\left(
           	1-\frac{4m_f^2}{m_{h}^2}
           	\right)^{3/2}
           	+
           	\left|
           	g^{A}_{h f\bar f}
           	\right|^2
           	\left(
           	1-\frac{4m_f^2}{m_{h}^2}
           	\right)^{3/2}
           	\right].
           	\label{eq:Gamma_hi_ff}
           \end{align}
           Here, $G_F$ is the Fermi constant, $g^S$ denotes the CP-even scalar Higgs coupling, and $g^A$ denotes the CP-odd scalar Higgs coupling. Moreover, $N_c$ is the color factor, with $N_c=3$ for quarks, $N_c=1$ for leptons, and $e_q$ denotes the electric charge of the quark $q$ in units of the elementary charge $e$. For completeness, lengthy analytic expressions and additional technical details are deferred to the Appendix~\ref{app:loopfuctions}.
           
           To quantify the overall consistency between the model predictions and the measured properties of the $125~\mathrm{GeV}$ Higgs boson, we construct a combined $\chi^2$ function, which includes the mass of the SM-like state $h$ and the five signal strengths introduced above
           \begin{align}
           	\chi^2_{h}
           	=
           	\left(
           	\frac{m_{h}^{\mathrm{th}}-m_{h}^{\mathrm{exp}}}{\delta_{m_h}}
           	\right)^{2}
           	+
           	\sum_{X}
           	\left(
           	\frac{\mu_{X}^{\mathrm{th}}-\mu_{X}^{\mathrm{exp}}}{\delta_{X}}
           	\right)^{2},
           	\qquad
           	X\in\{\gamma\gamma,\,ZZ^{*},\,WW^{*},\,b\bar b,\,\tau^{+}\tau^{-}\}.
           	\label{eq:chi2_higgs}
           \end{align}
           Here, $\mu_{X}^{\mathrm{th}}$ denotes the predicted signal strength obtained from eqs.~(\ref{eq:Gamma_hi_gg})--(\ref{eq:Gamma_hi_ff}), while $\mu_{X}^{\mathrm{exp}}\pm\delta_{X}$ are the corresponding experimental central values and uncertainties listed in eqs.~(\ref{eq:mu_h2_gammagamma})--(\ref{eq:mu_h2_tautau}). A parameter point is regarded as compatible with the Higgs data if it lies within the $2\sigma$, namely $95.45\%$ confidence-level, region of the combined fit.
           
           \subsection{Electroweak precision constraints}
           
           The electroweak precision selection uses
           \begin{equation}
           	S=-0.04\pm0.10,\qquad T=0.01\pm0.12,\qquad U=-0.01\pm0.11.
           	\label{eq:STU-constraints}
           \end{equation}
           The constraints on $S$, $T$, and $U$ are imposed as mutually independent $1\sigma$ intervals and are used as a simplified electroweak precision selection criterion rather than a full correlated electroweak likelihood.

           \section{Numerical analysis}
           \label{sec:Nuanalysis}
           
           In this section, we present the numerical results for the $h \to bs$. For relevant parameters in the SM, we choose  
           \begin{equation}
           	\begin{aligned}[b]
           		& \alpha_s(m_Z) = 0.118,\quad m_t = 163\,\mathrm{GeV},\quad m_c = 0.64\,\mathrm{GeV},\quad m_Z = 91.1876\,\mathrm{GeV},\\
           		& \alpha(m_Z)   = \frac{1}{128.9},\quad m_b = 2.84\,\mathrm{GeV},\quad m_s = 0.095\,\mathrm{GeV},\quad m_W = 80.385\,\mathrm{GeV}.
           		\label{eq:4}
           	\end{aligned}
           \end{equation}
           here, \(\alpha_s(m_Z)\) and \(\alpha(m_Z)\) represent the strong coupling constant and the electromagnetic fine-structure constant, respectively, evaluated at the scale \(m_Z\).

           \begin{table}[htbp]
           	\centering
           	\begin{tabular*}{0.9\textwidth}{@{\extracolsep{\fill}}lcc}
           		\hline
           		Parameters & Min & Max \\
           		\hline
           		$\lambda_3$ & 0 & 1 \\
           		$\lambda_i\ (i \neq 3)$ & 0 & 4 \\
           		$\tan\beta$ & 3 & 10 \\
           		$\tan\beta'$ & 15 & 25 \\
           		$M_{H_{min}^\pm}~(\mathrm{GeV})$ & 300 & 1500 \\
           		$M_{H_{max}^\pm}~(\mathrm{GeV})$ & 3000 & 5000 \\
           		$\gamma_{\pm}/\pi$ & -1 & 1 \\
           		\hline
           	\end{tabular*}
           	\caption{Parameter ranges used in the numerical analysis}
           	\label{tab:61}
           \end{table}
           where, $H_{\min}^{\pm}$ and $H_{\max}^{\pm}$ denote the lightest and heaviest charged Higgs bosons, respectively. On the one hand, nonzero additional Yukawa parameters $O^{l}_{ij}$ and $O^{u}_{ij}$ may introduce new tree-level FCNC in the charged-lepton and up-type-quark sectors. On the other hand, the parameters $O^{d}_{ij}$ in the down-type-quark sector are stringently constrained by neutral-meson mixing observables. To reduce the number of free parameters and highlight the effects of the scalar sector on the $h\to bs$ decay, we adopt the simplified setting $O^{q}_{ij}=0$, where $q=l,u,d$, and set the corresponding complex phases to zero.
           
           Then, we impose all the theoretical and experimental constraints and
           retain only the parameter points that satisfy them simultaneously.
           Based on the surviving samples, we investigate the dependence of
           $\mathrm{BR}_{\rm full}(h\to bs)$ on the model parameters. Since the
           G3HDM parameter space is multidimensional and the dependence of the
           branching ratio on the input parameters is not necessarily linear, we
           employ the Spearman rank-correlation coefficient to quantify their
           monotonic correlations.
           
           For two variables $X$ and $Y$, the Spearman rank-correlation
           coefficient is defined as
           \begin{equation}
           	\rho(X,Y)
           	=
           	\frac{
           		\operatorname{Cov}(R_X,R_Y)
           	}{
           		\sigma_{R_X}\sigma_{R_Y}
           	},
           	\label{eq:Spearman}
           \end{equation}
           where $R_X$ and $R_Y$ denote the ranks of $X$ and $Y$, respectively,
           and $\sigma_{R_X}$ and $\sigma_{R_Y}$ are their standard deviations.
           The coefficient satisfies $-1\leq\rho\leq1$. Values close
           to $+1$ and $-1$ indicate strong positive and negative monotonic
           correlations, respectively, whereas a value close to zero indicates
           a weak monotonic correlation.
           
\begin{figure}[htbp]
           	\centering
           	\includegraphics[width=0.7\textwidth]{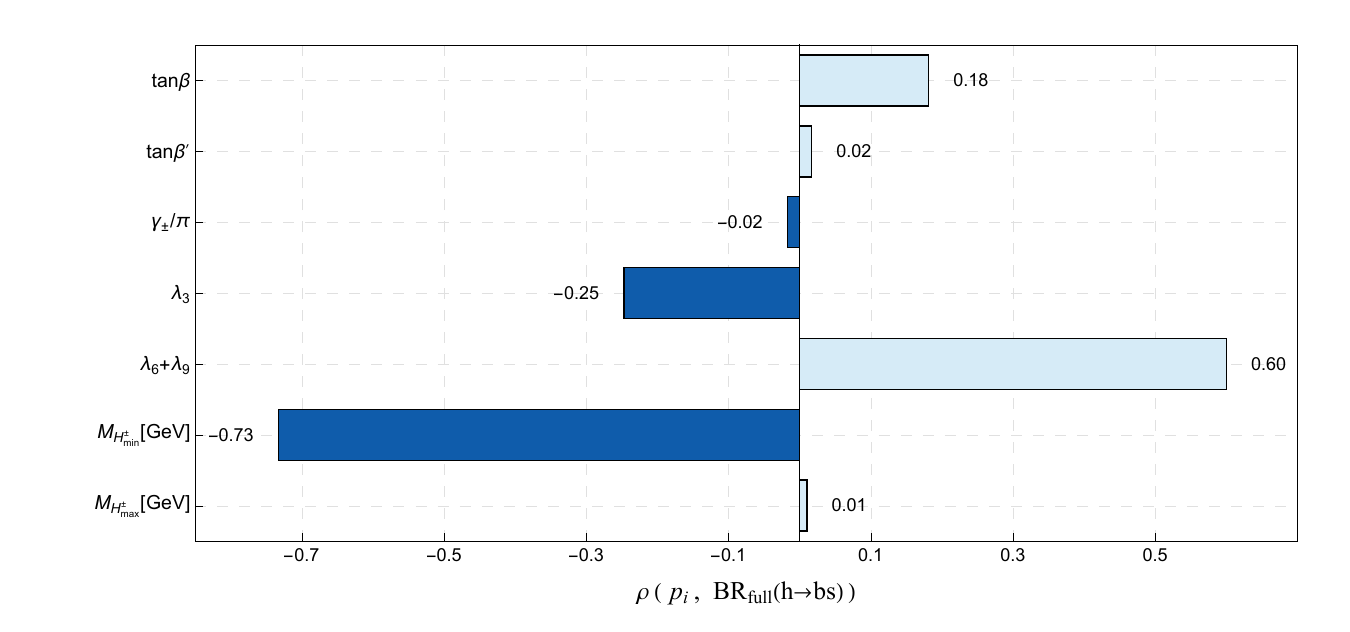}
           	\caption{\raggedright Spearman rank-correlation coefficients $\rho$ between $\mathrm{BR}_{\rm full}(h\to bs)$ and the selected scalar-sector parameters for the parameter points satisfying all the imposed theoretical and experimental constraints. The light and dark bars represent positive and negative correlations, respectively.}
           	\label{fig:Sperman picture}
           \end{figure}

           In the present analysis, we calculate
           $\rho\!\left(p_i,\mathrm{BR}_{\rm full}(h\to bs)\right)$
           for each scanned parameter $p_i$. The results are presented in
           Fig.~\ref{fig:Sperman picture}.
           
\begin{figure}[htbp]
           	\centering
           	\includegraphics[width=0.48\textwidth]{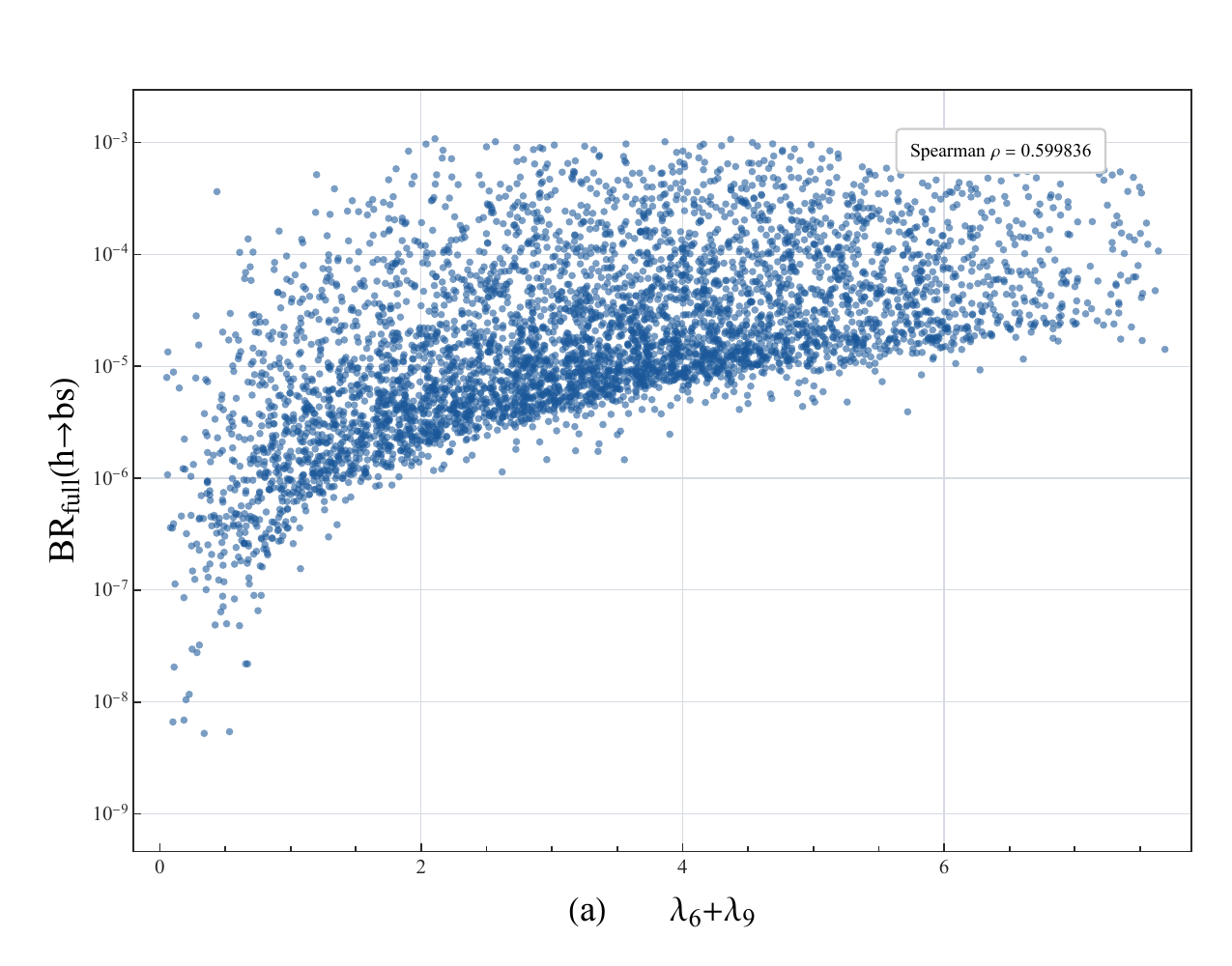}
           	\hfill
           	\includegraphics[width=0.48\textwidth]{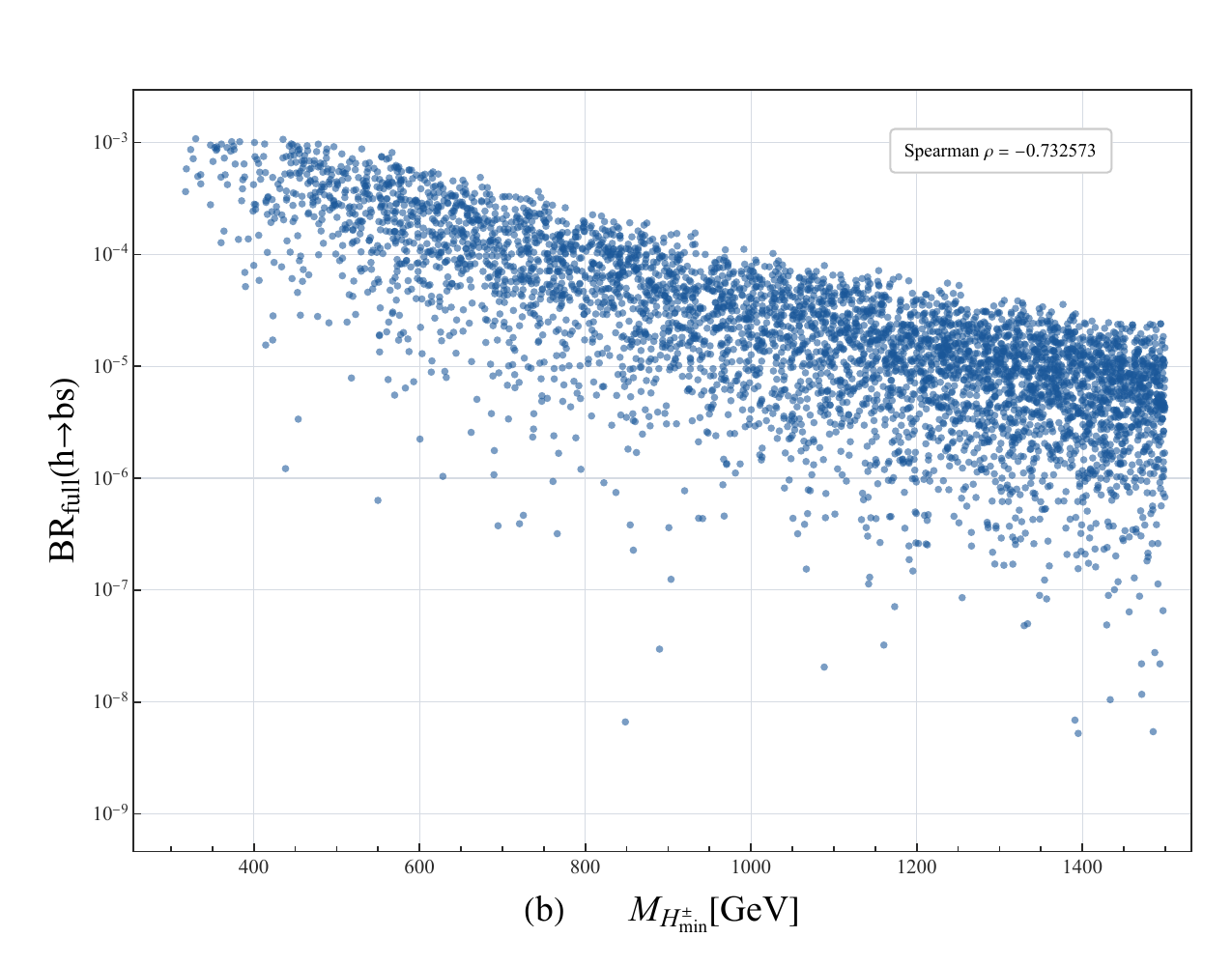}
           	\caption{\raggedright Dependence of $\mathrm{BR}_{\rm full}(h\to bs)$ on
           		(a) the scalar-potential combination $\lambda_6+\lambda_9$ and
           		(b) the lightest charged-Higgs mass $M_{H_{min}^\pm}$,
           		for the parameter points satisfying all the imposed theoretical and
           		experimental constraints. The corresponding Spearman rank-correlation
           		coefficients are $\rho=0.60$ and $-0.73$, respectively.}
           	\label{fig:ParameterBR}
           \end{figure}

           As shown in Fig.~\ref{fig:Sperman picture}, among the parameters considered, the scalar-potential parameter combination $\lambda_6+\lambda_9$ and the mass of the lightest charged Higgs boson, $M_{H_{\min}^{\pm}}$, both exhibit strong correlations with $\mathrm{BR}_{\rm full}(h\to bs)$. We therefore present the corresponding scatter plots in Fig.~\ref{fig:ParameterBR}, which show that the branching ratio can still reach the $10^{-3}$ level while satisfying all the imposed constraints.
           
           Figure~\ref{fig:ParameterBR}(a) shows that the upper envelope of the branching ratio generally increases with $\lambda_6+\lambda_9$, consistent with the positive Spearman correlation. In contrast, Fig.~\ref{fig:ParameterBR}(b) exhibits a clear negative correlation between the branching ratio and $M_{H_{\min}^{\pm}}$, indicating that large values of $\mathrm{BR}(h\to bs)$ preferentially occur for a relatively light charged-Higgs state. However, these correlations alone do not establish the dynamical origin of the large branching ratio, which will be further clarified below.

           To identify the dynamical origin of the large branching ratio, we decompose the full decay amplitude as follows
           \begin{equation}
           	\mathcal{M}_{\rm full}
           	=
           	\mathcal{M}_{\rm tree}
           	+
           	\mathcal{M}_{\rm neutral}
           	+
           	\mathcal{M}_{\rm charged}
           	+
           	\mathcal{M}_{W}.
           	\label{eq:amplitude-decomposition}
           \end{equation}
           where $\mathcal{M}_{\rm neutral}$, $\mathcal{M}_{\rm charged}$, and
           $\mathcal{M}_{W}$ denote the contributions from the neutral-Higgs,
           charged-Higgs, and $W$ sectors, respectively.
           
           For comparison, we define the individual branching ratios by retaining
           only the corresponding contribution in the decay amplitude. These
           quantities are introduced as diagnostic measures and are not additive,
           since the full branching ratio is proportional to
           $|\mathcal{M}_{\rm full}|^2$ and therefore contains interference terms.
           We first define the incoherent sum of the standalone contributions as
           \begin{equation}
           	\mathrm{BR}_{\rm inc}
           	=
           	\mathrm{BR}_{\rm tree}
           	+
           	\mathrm{BR}_{\rm neutral}
           	+
           	\mathrm{BR}_{\rm charged}
           	+
           	\mathrm{BR}_{W}.
           	\label{eq:incoherent-BR}
           \end{equation}
           The net interference contribution is then given by
           \begin{equation}
           	\Delta\mathrm{BR}_{\rm int}
           	=
           	\mathrm{BR}_{\rm full}
           	-
           	\mathrm{BR}_{\rm inc}.
           	\label{eq:interference-BR}
           \end{equation}
           To quantify its relative importance, we further introduce the
           normalized net interference
           \begin{equation}
           	\eta_{\rm int}
           	=
           	\frac{\Delta\mathrm{BR}_{\rm int}}
           	{\mathrm{BR}_{\rm inc}}
           	\times100\%.
           	\label{eq:normalized-interference}
           \end{equation}
           Accordingly, $\eta_{\rm int}>0$ corresponds to net constructive
           interference, whereas $\eta_{\rm int}<0$ indicates net destructive
           interference.
           
           \begin{figure}[htbp]
           	\centering
           	\includegraphics[width=0.48\textwidth]{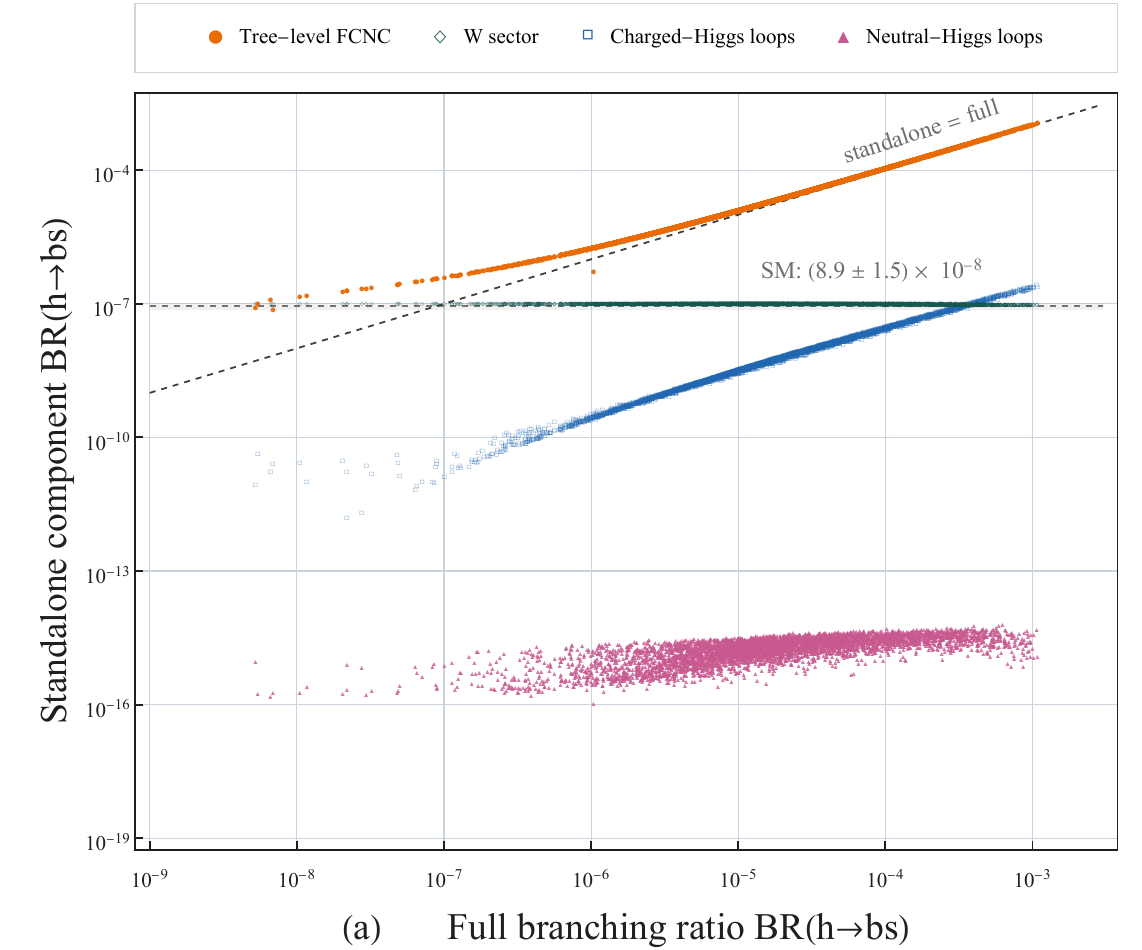}
           	\hfill
           	\includegraphics[width=0.48\textwidth]{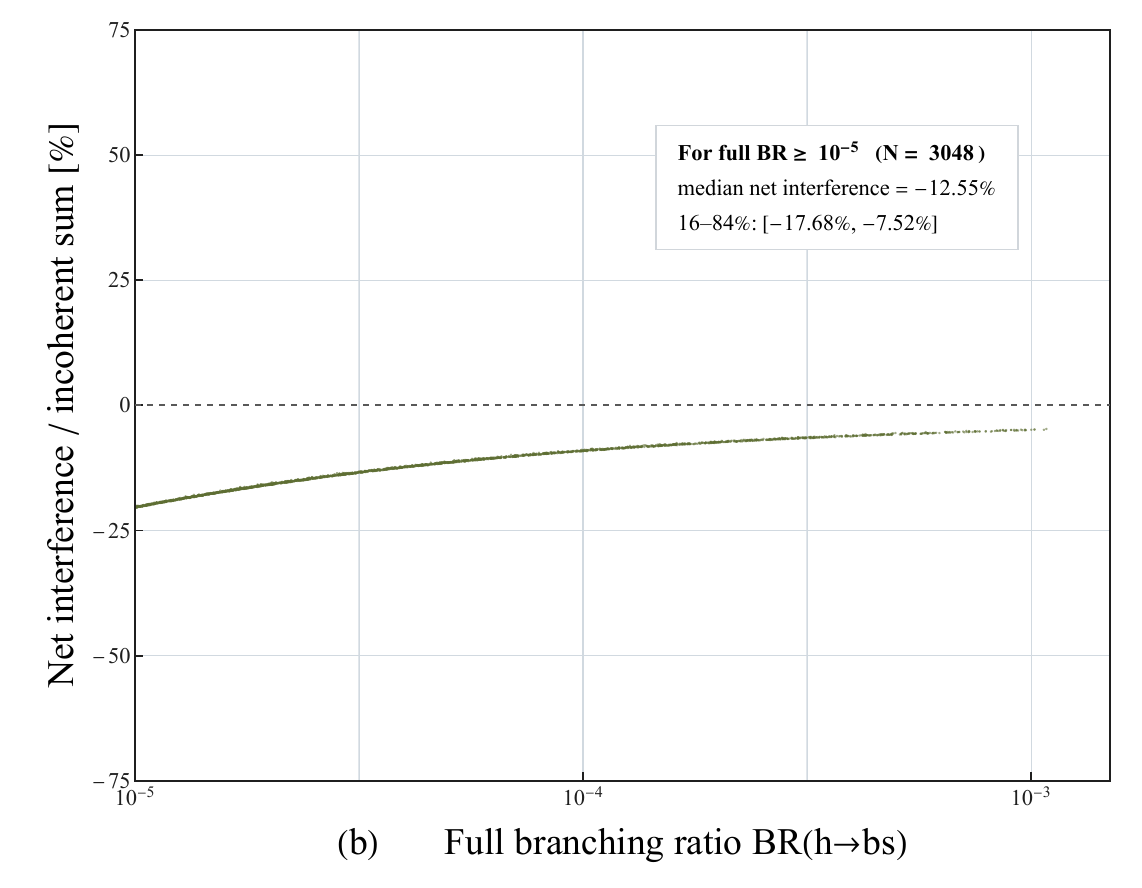}
           	\caption{\raggedright Panel (a) shows the standalone tree-level FCNC, $W$-sector, charged-Higgs-loop, and neutral-Higgs-loop contributions as functions of $\mathrm{BR}_{\rm full}(h\to bs)$. The diagonal dashed line represents the case in which a standalone contribution is equal to the full result, while the horizontal band denotes the SM prediction $\mathrm{BR}_{\rm SM}(h\to bs)=(8.9\pm1.5)\times10^{-8}$. Panel (b) shows the net interference as a function of $\mathrm{BR}_{\rm full}(h\to bs)$ for the 3048 parameter points satisfying $\mathrm{BR}_{\rm full}\geq10^{-5}$. The horizontal dashed line corresponds to zero net interference.}
           		\label{fig:AmplitudeDecomposition}
           \end{figure}
           
           The standalone contributions and the normalized interference are shown in Fig.~\ref{fig:AmplitudeDecomposition}. Figure~\ref{fig:AmplitudeDecomposition}(a) compares the tree-level and individual loop contributions with the full branching ratio, while Fig.~\ref{fig:AmplitudeDecomposition}(b) displays $\eta_{\rm int}$ for the large-branching-ratio region, $\mathrm{BR}_{\rm full}\geq10^{-5}$.
           
           As shown in Fig.~\ref{fig:AmplitudeDecomposition}(a), in the region with a relatively large total branching ratio, the tree-level FCNC contribution closely follows the diagonal line $\mathrm{BR}_{\rm tree}=\mathrm{BR}_{\rm full}$. By contrast, the contributions from the $W$ sector, charged-Higgs loops, and neutral-Higgs loops are all significantly smaller. Therefore, the parameter points satisfying $\mathrm{BR}_{\rm full}(h\to bs)\sim10^{-3}$ are predominantly dominated by the tree-level FCNC amplitude.
           
           Figure~\ref{fig:AmplitudeDecomposition}(b) shows that the interference is predominantly destructive in the region $\mathrm{BR}_{\rm full}\geq10^{-5}$. The median normalized interference is $-12.55\%$, with a $16$--$84\%$ interval of $[-17.68\%,-7.52\%]$. Thus, interference typically reduces the incoherent result by approximately $10$--$20\%$, but does not alter the conclusion that the large branching ratio is predominantly dominated by the tree-level FCNC contribution.

           Since the tree-level amplitude is directly determined by the FCNC Yukawa vertex $h\bar bs$, the problem can be further reduced to an analysis of the corresponding effective FCNC coupling. Given that $m_b,m_s\ll m_h$, the interference term between the left and right handed chiral structures is suppressed by the quark masses. Therefore, the tree-level decay rate is governed predominantly by the following quadratic combination
           \begin{equation}
           	|Y_{\rm FCNC}^{h}|^2
           	\equiv
           	|C_{h\bar b s}^{L}|^2
           	+
           	|C_{h\bar b s}^{R}|^2.
           	\label{eq:YFCNC-definition}
           \end{equation}
           Here, $C_{h\bar b s}^{L}$ and $C_{h\bar b s}^{R}$ denote the coefficients multiplying the left  and right handed chiral projectors, respectively, in the $h\bar b s$ Yukawa vertex. Then, the tree-level branching ratio then satisfies
           \begin{equation}
           	\mathrm{BR}_{\rm tree}(h\to bs)
           	\propto
           	|Y_{\rm FCNC}^{h}|^2,
           	\label{eq:BRtree-YFCNC}
           \end{equation}
           
           Under the Yukawa-input convention adopted in our numerical analysis, $C_{h\bar b s}^{R}$ vanishes, while $C_{h\bar b s}^{L}$ remains nonzero because of the irreducible CKM-induced structure of the down-type-quark mass matrices. Nevertheless, both chiral couplings are retained in Eq.~(\ref{eq:YFCNC-definition}) for completeness.
          
           \begin{figure}[htbp]
           	\centering
           	\includegraphics[width=0.48\textwidth]{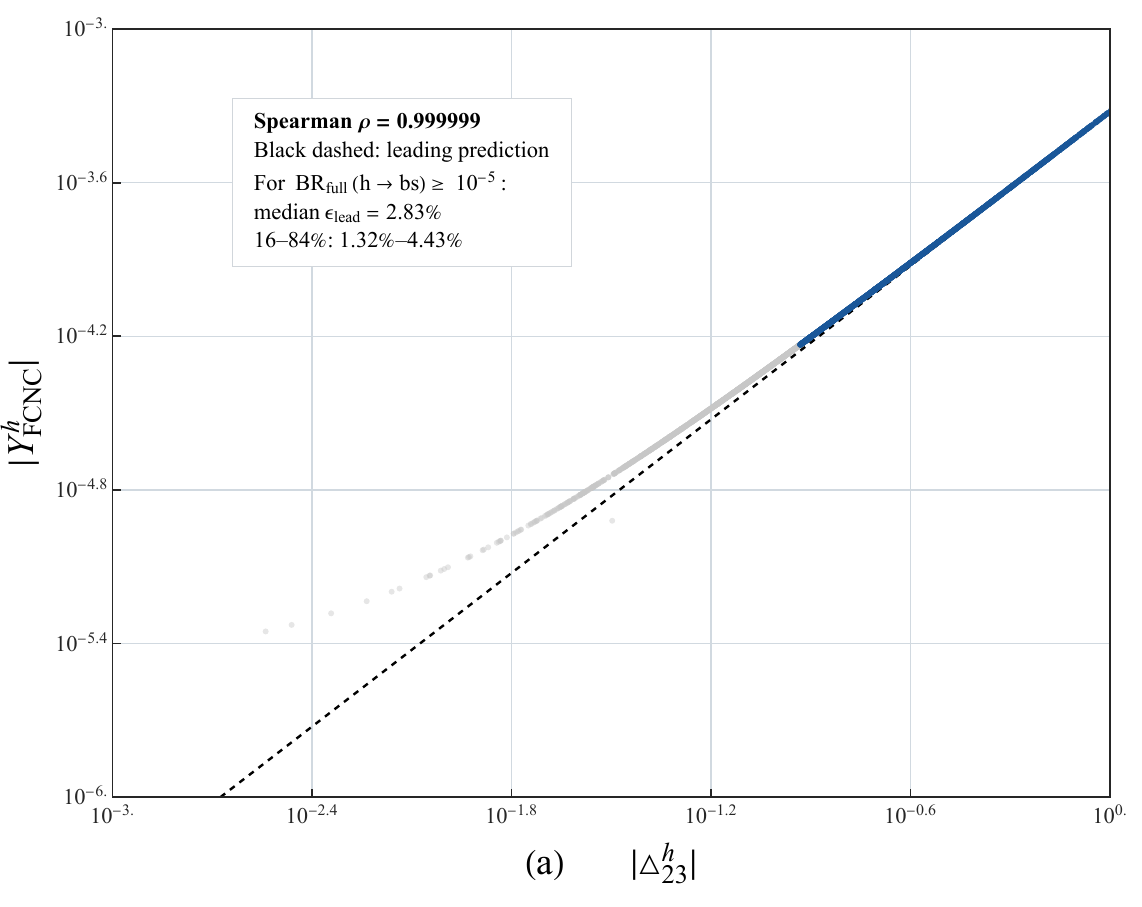}
           	\hfill
           	\includegraphics[width=0.48\textwidth]{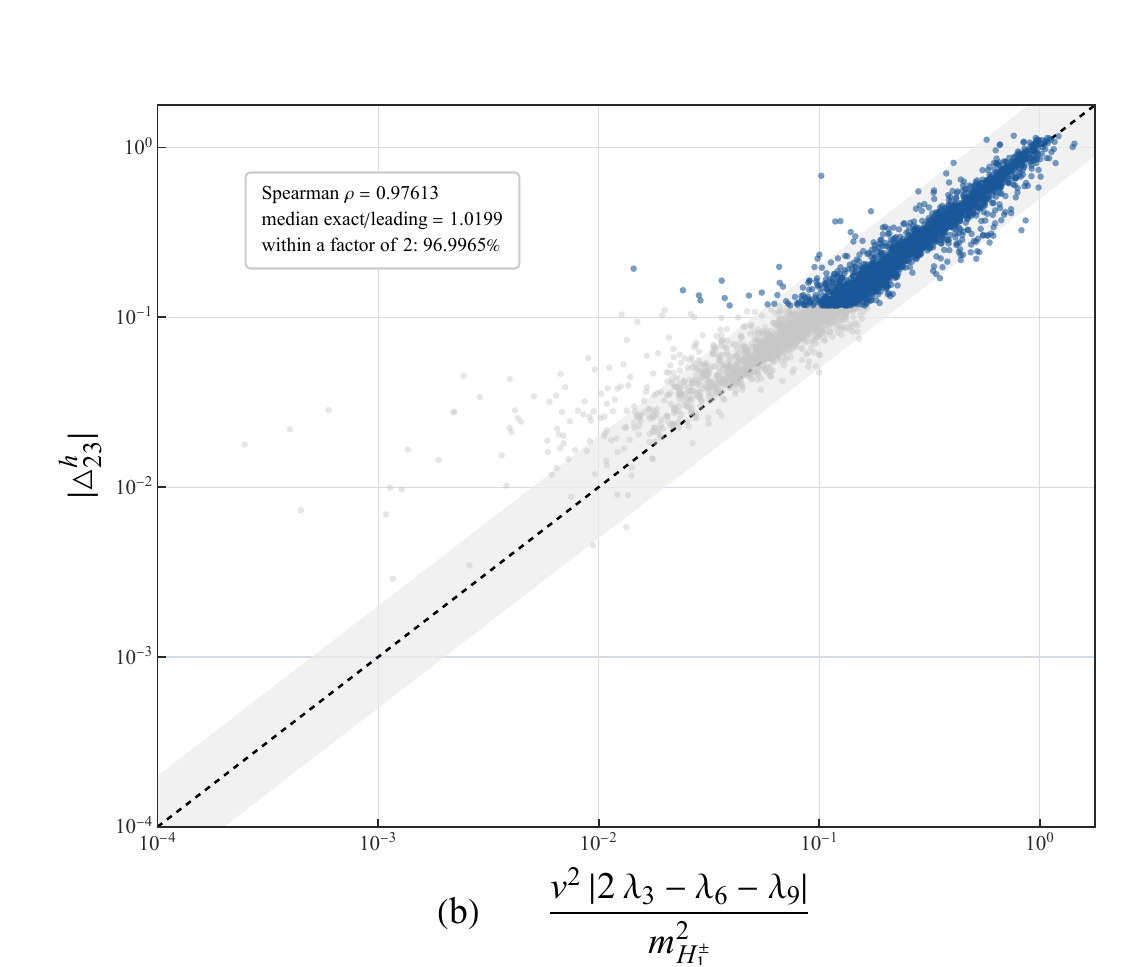}
           	\caption{\raggedright Panel (a) shows the effective FCNC coupling strength $|Y_{\rm FCNC}^{h}|$ as a function of the scalar-misalignment parameter $|\Delta_{23}^{h}|$. The blue and light-gray points correspond to $\mathrm{BR}_{\rm full}(h\to bs)\geq 10^{-5}$ and $\mathrm{BR}_{\rm full}(h\to bs)<10^{-5}$, respectively. The black dashed line represents the leading theoretical relation $|Y_{\rm FCNC}^{h}| =(m_b/v)|V_{cs}V_{cb}^{*}|\,|\Delta_{23}^{h}|$ and is not fitted to the numerical samples. Panel (b) compares the exact value of $|\Delta_{23}^{h}|$ with its leading scalar-sector estimate $v^{2}|2\lambda_{3}-\lambda_{6}-\lambda_{9}| /M_{H_{min}^\pm}^{2}$, where the approximate relation $M_{A_{min}}\simeq M_{H_{min}^\pm}$ in the surviving parameter space has been used. The black dashed line denotes equality between the exact and leading results, while the gray band indicates agreement within a factor of two. The same color convention as in panel (a) is adopted.
           	}
           	\label{fig:Delta23Origin}
           \end{figure}
           
           Therefore, the nonvanishing chiral coupling can be written as
           \begin{equation}
           	C_{h\bar b s}^{L}
           	=
           	-m_b\left(
           	\frac{Z_{hH}^H}{v_2}V_{cs}V_{cb}^{*}
           	+
           	\frac{Z_{hH^{\prime}}^H}{v_3}V_{ts}V_{tb}^{*}
           	\right).
           	\label{eq:CL-texture}
           \end{equation}
           Using the CKM unitarity relation $V_{us}V_{ub}^{*}+V_{cs}V_{cb}^{*}+V_{ts}V_{tb}^{*}=0$, this coupling becomes
           \begin{equation}
           	C_{h\bar b s}^{L}
           	=
           	-\frac{m_b}{v}V_{cs}V_{cb}^{*}\Delta_{23}^{h}
           	+
           	C_{\rm res}^{L},
           	\qquad
           	\Delta_{23}^{h}
           	\equiv
           	v\left(
           	\frac{Z_{hH}^H}{v_2}
           	-
           	\frac{Z_{hH^{\prime}}^H}{v_3}
           	\right),
           	\label{eq:CL-Delta23}
           \end{equation}
           where $C_{\rm res}^{L}=m_bV_{us}V_{ub}Z_{hH^{\prime}}^H/v_3$. Since $|V_{us}V_{ub}|\ll|V_{cs}V_{cb}^{*}|$, the residual term is CKM suppressed, and the dominant part of the FCNC coupling is therefore controlled by $\Delta_{23}^{h}$.

           In the exact alignment limit, the light Higgs boson is aligned with the vacuum direction, so that $Z_{hi}^H=v_i/v$. Consequently, $Z_{hH}^H/v_2=Z_{hH^{\prime}}^H/v_3=1/v$ and $\Delta_{23}^{h}=0$. Thus, $\Delta_{23}^{h}$ directly measures the scalar misalignment responsible for the dominant tree-level FCNC coupling.
           
           Using Eq.~(\ref{eq:YFCNC-definition}) and Eq.~(\ref{eq:CL-Delta23}), the effective FCNC coupling satisfies
           \begin{equation}
           	|Y_{\rm FCNC}^{h}|
           	=
           	\left|
           	-\frac{m_b}{v}V_{cs}V_{cb}^{*}\Delta_{23}^{h}
           	+
           	C_{\rm res}^{L}
           	\right|.
           	\label{eq:YFCNC-Delta23-exact}
           \end{equation}
           The leading contribution is
           \begin{equation}
           	|Y_{\rm FCNC}^{h}|
           	\simeq
           	|Y_{\Delta_{23}}^{h}|
           	\equiv
           	\frac{m_b}{v}
           	|V_{cs}V_{cb}^{*}|
           	|\Delta_{23}^{h}|.
           	\label{eq:YFCNC-Delta23-leading}
           \end{equation}
           
           To numerically test the above leading-order relation, Fig.~\ref{fig:Delta23Origin}(a) shows $|Y_{\rm FCNC}^{h}|$ as a function of $|\Delta_{23}^{h}|$. The gray points represent the parameter points satisfying $\mathrm{BR}_{\rm full}(h\to bs)<10^{-5}$, whereas the blue points correspond to $\mathrm{BR}_{\rm full}(h\to bs)\geq10^{-5}$. The black dashed line represents the theoretical prediction given by Eq.~(\ref{eq:YFCNC-Delta23-leading}). The nearly perfect Spearman correlation coefficient, $\rho=0.999999$, confirms that $|Y_{\rm FCNC}^{h}|$ is predominantly controlled by $|\Delta_{23}^{h}|$.

           For the large branching ratio region $\mathrm{BR}_{\rm full}(h\to bs)\geq10^{-5}$, we define
           \begin{equation}
           	\epsilon_{\rm lead}
           	=
           	\frac{
           		\left|
           		|Y_{\rm FCNC}^{h}|-|Y_{\Delta_{23}}^{h}|
           		\right|
           	}{
           		|Y_{\rm FCNC}^{h}|
           	}.
           	\label{eq:epsilon-leading}
           \end{equation}
           
           Its median value is only $2.83\%$, with a $16$--$84\%$ interval of $1.32\%$--$4.43\%$. Therefore, this leading-order relation provides an excellent description in the large branching ratio region.
           
           To understand the scalar-sector origin of $\Delta_{23}^{h}$, we consider the leading departure from the decoupling limit. The diagonalization matrices of the CP-even and CP-odd scalar sectors can be related by
           
           \begin{equation}
           	Z^H\simeq(\mathbf{1}+\delta)Z^A,
           	\qquad
           	\delta^{T}=-\delta,
           	\label{eq:OH-OA-smallrotation}
           \end{equation}
           where the matrix $\delta = Z^H Z^{AT}-\mathbf{1}$ captures the departure from the decoupling limit. We find
           \begin{align}
           	\delta_{12}
           	&=
           	\frac{v^{2}}{M_{A_{\min}}^{2}}
           	\frac{1}{\tan\beta}
           	\left(
           	2\lambda_{3}-\lambda_{6}-\lambda_{9}
           	\right)
           	=
           	\frac{v^{2}}{M_{A_{\min}}^{2}}
           	\frac{1}{\tan\beta}
           	\lambda_{23},
           	\label{eq:delta12}
           	\\
           	\delta_{13}
           	&=
           	\frac{v^{2}}{M_{A_{\max}}^{2}}
           	\frac{1}{\tan\beta'}
           	\left(
           	2\lambda_{3}-\lambda_{5}-\lambda_{8}
           	\right)
           	=
           	\frac{v^{2}}{M_{A_{\max}}^{2}}
           	\frac{1}{\tan\beta'}
           	\lambda_{13},
           	\label{eq:delta13}
           	\\
           	\delta_{23}
           	&=
           	-\frac{v^{2}}{M_{A_{\max}}^{2}}
           	\frac{1}{\tan\beta\,\tan\beta'}
           	\left(
           	2\lambda_{3}+\lambda_{4}-\lambda_{5}-\lambda_{6}
           	+\lambda_{7}-\lambda_{8}-\lambda_{9}
           	\right)
           	=
           	-\frac{v^{2}}{M_{A_{\max}}^{2}}
           	\frac{1}{\tan\beta\,\tan\beta'}
           	\lambda_{12}.
           	\label{eq:delta23}
           \end{align}
           Taking into account that the mixing angle $\gamma_A$ in the pseudoscalar sector is small, with an order of magnitude given by $\gamma_A\sim 1/\tan\beta'\ll 1$, therefore
           \begin{align}
           	Z^{A}
           	&\simeq
           	\begin{pmatrix}
           		\sin\beta' & 0 & -\cos\beta' \\
           		0 & 1 & 0 \\
           		\cos\beta' & 0 & \sin\beta'
           	\end{pmatrix}
           	\begin{pmatrix}
           		1 & 0 & 0 \\
           		0 & \sin\beta & -\cos\beta \\
           		0 & \cos\beta & \sin\beta
           	\end{pmatrix}.
           	\label{eq:ZA}
           \end{align}
           Using Eq.~(\ref{eq:delta12}) and Eq.~(\ref{eq:ZA}), we obtain
           \begin{equation}
           	|\Delta_{23}^{h}|
           	\simeq
           	\frac{
           		v^{2}|2\lambda_{3}-\lambda_{6}-\lambda_{9}|
           	}{
           		M_{A_{min}}^{2}
           	}
           	\simeq
           	\frac{
           		v^{2}|2\lambda_{3}-\lambda_{6}-\lambda_{9}|
           	}{
           		M_{H_{min}^\pm}^{2}
           	},
           	\label{eq:Delta23-leading}
           \end{equation}
           where the last approximation follows from $M_{A_{min}}\simeq M_{H_{min}^\pm}$ in the surviving parameter space.
           
           Eq.~(\ref{eq:Delta23-leading}) explains the strong correlation observed in Fig.~\ref{fig:Delta23Origin}(b), while the remaining spread around the line of equality arises from finite-mixing effects and the influence of the other parameters in the full CP-even scalar mass matrix.
           
           Combining the results of the two panels, the enhancement mechanism can be summarized as
           \begin{equation}
           	\frac{v^{2}}{M_{H_{min}^\pm}^{2}}
           	\left|
           	2\lambda_3-(\lambda_6+\lambda_9)
           	\right|
           	\longrightarrow
           	|\Delta_{23}^{h}|
           	\longrightarrow
           	|Y_{\rm FCNC}^{h}|
           	\longrightarrow
           	\mathrm{BR}_{\rm tree}(h\to bs),
           	\label{eq:hbs-enhancement-chain}
           \end{equation}
           
           In the hierarchical-VEV and decoupling limits, the mass of the SM-like Higgs boson satisfies
           \begin{equation}
           	M_h^2\simeq 2v^2\lambda_3.
           	\label{eq:mh-lambda3}
           \end{equation}
           Therefore, requiring $M_h$ to remain close to $125.2~\mathrm{GeV}$ restricts $\lambda_3$ to a comparatively narrow range around $\lambda_3\simeq M_h^2/(2v^2)$. Consequently, although $\lambda_3$ appears explicitly in Eq.~(\ref{eq:hbs-enhancement-chain}), its limited range of variation weakens its Spearman correlation with $\mathrm{BR}_{\rm full}(h\to bs)$.
           
           The variation of $|2\lambda_3-\lambda_6-\lambda_9|$ over the surviving parameter space is therefore driven mainly by $\lambda_6+\lambda_9$, explaining the correlation observed in Fig.~\ref{fig:ParameterBR}(a). At the same time, the inverse dependence on $M_{H_{min}^\pm}^{2}$ accounts for the negative correlation shown in Fig.~\ref{fig:ParameterBR}(b). A sizable scalar-potential combination together with a relatively light scalar mass scale enhances $|\Delta_{23}^{h}|$, which in turn increases the tree-level FCNC Yukawa coupling and allows $\mathrm{BR}_{\rm full}(h\to bs)$ to reach the order of $10^{-3}$ in the allowed parameter space.
           
           We emphasize that the strong correlation with $M_{H_{\min}^{\pm}}$ does not imply that charged-Higgs loop diagrams dominate this decay. Instead, $M_{H_{\min}^{\pm}}\simeq M_{A_{\min}}$ controls the scalar-misalignment parameter through its appearance in the denominator, thereby affecting the tree-level FCNC coupling.

           Having identified the dynamical origin of the large branching ratio, we finally examine the extent to which the surviving parameter space can be tested by future collider experiments. In this work, we map several representative projected upper limits on $\mathrm{BR}(h\to bs)$ onto the accepted parameter sample.
           
           For a reference projected upper limit $\mathrm{BR}_{\rm sens}$, we define the model-space coverage fraction as
           \begin{equation}
           	f_{\rm cov}(\mathrm{BR}_{\rm sens})
           	=
           	\frac{
           		N\!\left[
           		\mathrm{BR}_{\rm full}(h\to bs)
           		\geq \mathrm{BR}_{\rm sens}
           		\right]
           	}{
           		N_{\rm acc}
           	},
           	\label{eq:coverage-fraction}
           \end{equation}
           where $N_{\rm acc}=4861$ is the number of parameter points that satisfy all the theoretical and experimental constraints. Therefore, $f_{\rm cov}$ represents the fraction of the accepted parameter space that could be tested by an experiment reaching the corresponding projected upper limit. Accordingly, a smaller value of $\mathrm{BR}_{\rm sens}$ corresponds to a higher experimental sensitivity.
           
           \begin{figure}[htbp]
           	\centering
           	\includegraphics[width=0.7\textwidth]{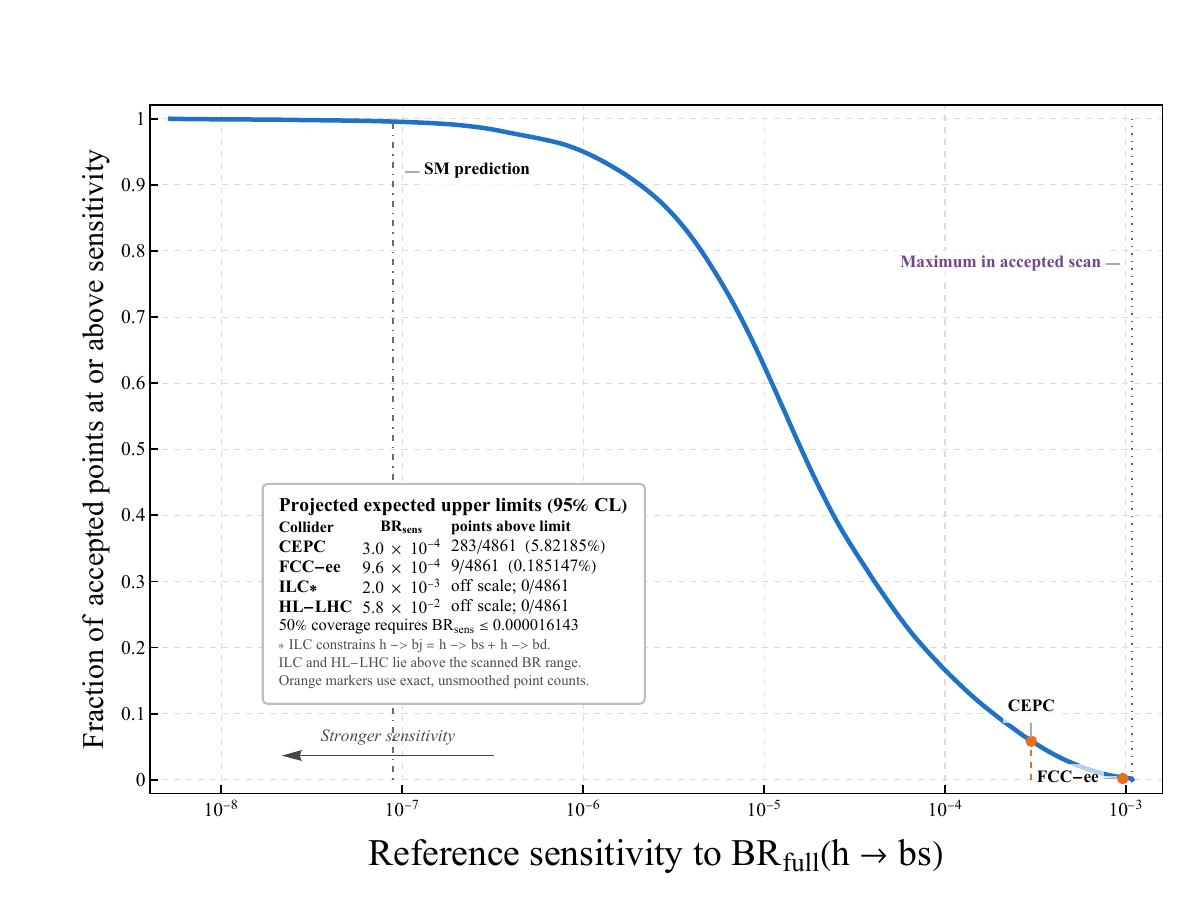}
           	\caption{\raggedright Fraction of the accepted parameter points satisfying $\mathrm{BR}_{\rm full}(h\to bs)\geq\mathrm{BR}_{\rm sens}$ as a function of the reference branching-ratio sensitivity $\mathrm{BR}_{\rm sens}$. The blue curve shows the model-space coverage fraction, while the orange markers denote the representative CEPC and FCC-ee sensitivities. The vertical lines indicate the SM prediction and the maximum branching ratio obtained in the accepted scan. Smaller values of $\mathrm{BR}_{\rm sens}$ correspond to stronger experimental sensitivity.}
           	\label{fig:FutureColliderCoverage}
           \end{figure}
           
           Figure~\ref{fig:FutureColliderCoverage} shows $f_{\rm cov}$ as a function of $\mathrm{BR}_{\rm sens}$. The blue curve represents the cumulative distribution of the branching ratios predicted by the accepted parameter points, while the orange markers indicate the coverage fractions corresponding to the representative CEPC and FCC-ee sensitivities. The vertical reference lines denote the SM prediction, $\mathrm{BR}_{\rm SM}(h\to bs)=8.9\times10^{-8}$, and the maximum branching ratio obtained in the accepted scan.
           
           Near the SM prediction, the coverage fraction is close to $1$, indicating that almost all accepted parameter points predict branching ratios above the SM expectation. In contrast, when the experimental sensitivity is of order $10^{-4}$--$10^{-3}$, only the tail of the parameter space with relatively large branching ratios can be tested.
           
           For the representative CEPC projected upper limit $\mathrm{BR}_{\rm sens}=3.0\times10^{-4}$, 283 of the 4861 accepted parameter points lie above the corresponding sensitivity, yielding $f_{\rm cov}=5.82\%$. For FCC-ee, we adopt $\mathrm{BR}_{\rm sens}=9.6\times10^{-4}$, which probes 9 accepted parameter points and corresponds to $f_{\rm cov}=0.185\%$.
           
           The median of the accepted branching-ratio distribution is $1.61\times10^{-5}$. Therefore, testing at least $50\%$ of the allowed parameter space would require a sensitivity of approximately $\mathrm{BR}_{\rm sens}\lesssim1.61\times10^{-5}$. This result shows that the representative CEPC and FCC-ee sensitivities can probe the region with the most significant branching-ratio enhancement, including part of the parameter space in which the decay is dominated by the tree-level FCNC interaction. However, a sensitivity approaching the $10^{-5}$ level would be required to probe a substantial fraction of the surviving G3HDM parameter space.
           
           We emphasize that Fig.~\ref{fig:FutureColliderCoverage} provides an indicative estimate of the model-space coverage rather than a detector-level projection. The actual experimental reach will also depend on the Higgs production channels, integrated luminosity, flavor-tagging efficiencies, and background-rejection capabilities. Nevertheless, the result demonstrates that the tree-level enhancement of $h\to bs$ in the G3HDM can bring part of the allowed parameter space within the reach of future collider experiments.

           \section{Conclusion}
           \label{sec:Conclusion}
           
           Within the framework of the G3HDM, we have systematically investigated the phenomenological properties of the flavor-changing Higgs decay $h\to bs$. In the SM, this process is strongly suppressed by the loop factor, the CKM matrix elements, and the GIM mechanism, leading to $\mathrm{BR}_{\rm SM}(h\to bs)\simeq 8.9\times10^{-8}$. Therefore, an observation of this decay with a significantly larger branching ratio would provide a sensitive probe of physics beyond the SM.
           
           In the numerical analysis, we impose constraints from the Higgs-boson signal strengths, electroweak precision observables, neutral-meson mixing, and the rare decays $\bar B\to X_s\gamma$ and $B_s^0\to\mu^+\mu^-$. With all these constraints imposed simultaneously, the branching ratio can still reach
           \begin{equation}
           	\mathrm{BR}_{\rm full}(h\to bs)\sim10^{-3},
           \end{equation}
           which is approximately four orders of magnitude larger than the SM prediction. The rank-correlation analysis further shows that large branching ratios preferentially occur for a relatively large $\lambda_6+\lambda_9$ and a relatively light charged-Higgs state.
           
           By decomposing the decay amplitude into the tree-level, neutral-scalar, charged-scalar, and $W$-boson loop contributions, we find that the large branching ratio is dominantly generated by the tree-level FCNC interaction rather than by the charged-Higgs loop diagrams. The corresponding FCNC Yukawa coupling is controlled by the scalar misalignment parameter
           \begin{equation}
           	\Delta_{23}^{h}
           	=
           	v\left(
           	\frac{Z^H_{hH}}{v_2}
           	-
           	\frac{Z^H_{hH^{\prime}}}{v_3}
           	\right).
           \end{equation}
           In the hierarchical-VEV and decoupling regime, its leading scalar-sector dependence is given by
           \begin{equation}
           	|\Delta_{23}^{h}|
           	\simeq
           	\frac{
           		v^2|2\lambda_3-\lambda_6-\lambda_9|
           	}{
           		M_{A_{\min}}^2
           	}
           	\simeq
           	\frac{
           		v^2|2\lambda_3-\lambda_6-\lambda_9|
           	}{
           		M_{H_{\min}^{\pm}}^{\,2}
           	}.
           \end{equation}
           The enhancement mechanism can therefore be summarized as
           \begin{equation}
           	\left\{
           	|2\lambda_3-\lambda_6-\lambda_9|,
           	M_{H_{\min}^{\pm}}
           	\right\}
           	\longrightarrow
           	|\Delta_{23}^{h}|
           	\longrightarrow
           	|Y_{\rm FCNC}^{h}|
           	\longrightarrow
           	\mathrm{BR}_{\rm tree}(h\to bs).
           \end{equation}
           This also clarifies that the strong correlation between $\mathrm{BR}_{\rm full}(h\to bs)$ and $M_{H_{\min}^{\pm}}$ originates mainly from the scalar-misalignment mass scale and does not imply that the charged-Higgs loop contribution dominates the decay.
           
           Our results demonstrate that the current low-energy flavor and electroweak constraints do not completely exclude an observable $h\to bs$ signal in the G3HDM. A portion of the surviving parameter space may be accessible to future high-precision Higgs measurements. Consequently, the decay $h\to bs$ provides a complementary probe of tree-level flavor-changing interactions and the extended Higgs sector of the G3HDM.

          \section*{Data Availability Statement}
          
          The numerical data generated and analyzed during the present study are available from the corresponding author upon reasonable request. No additional publicly available research data or software were generated beyond those presented in this article.

           \begin{acknowledgments}
           	The work has been supported by the National Natural Science Foundation of China
           	(NNSFC) with Grants No. 12075074, No. 12235008, No. 11535002, No. 11705045, Natural Science Foundation for Distinguished Young Scholars of Hebei Province with Grant No. A2022201017, Natural Science Foundation of Guangxi Autonomous Region with Grant No. 2022GXNSFDA035068, the youth top-notch talent support program of the Hebei Province, and Midwest Universities Comprehensive Strength Promotion project.
           \end{acknowledgments}
           
           \section*{Appendix: Loop functions}
           \label{app:loopfuctions}
            \begin{align}
            	I_1(x_1,x_2)
            	&= \frac{1+\ln x_2}{x_2-x_1}
            	+\frac{x_1\ln x_1-x_2\ln x_2}{(x_2-x_1)^2},
            	\notag \\[6pt]
            	I_2(x_1,x_2)
            	&= \frac{-1-\ln x_1}{x_2-x_1}
            	-\frac{x_1\ln x_1-x_2\ln x_2}{(x_2-x_1)^2},
            	\notag \\[6pt]
            	I_3(x_1,x_2)
            	&= \frac{1}{2}\left[
            	\frac{3+2\ln x_2}{x_2-x_1}
            	+\frac{-2x_2-4x_2\ln x_2}{(x_2-x_1)^2}
            	-\frac{2x_1^2\ln x_1}{(x_2-x_1)^3}
            	+\frac{2x_2^2\ln x_2}{(x_2-x_1)^3}
            	\right],
            	\notag \\[6pt]
            	I_4(x_1,x_2)
            	&= \frac{1}{4}\left[
            	\frac{11+6\ln x_2}{x_2-x_1}
            	+\frac{-15x_2-18x_2\ln x_2}{(x_2-x_1)^2}
            	+\frac{6x_2^2+18x_2^2\ln x_2}{(x_2-x_1)^3}
            	+\frac{6x_1^3\ln x_1-6x_2^3\ln x_2}{(x_2-x_1)^4}
            	\right].
            	\label{eq:loop-functions}
            \end{align}
           
           \begin{equation}
           	\label{eq:higgs-loop-functions}
           	\begin{aligned}[b]
           		&A_{1/2}(x)
           		=
           		\frac{2}{x^2}
           		\left[
           		x+(x-1)g(x)
           		\right],
           		\\[2mm]
           		&A_0(x)
           		=
           		-\frac{x-g(x)}{x^2},
           		\\[2mm]
           		&A_1(x)
           		=
           		-\frac{1}{x^2}
           		\left[
           		2x^2+3x+3(2x-1)g(x)
           		\right],
           		\\[2mm]
           		&A_2(x)
           		=
           		\frac{2g(x)}{x},
           		\\[2mm]
           		&g(x)
           		=
           		\begin{cases}
           			\arcsin^2\sqrt{x},
           			& x\leq 1,
           			\\[2mm]
           			-\dfrac{1}{4}
           			\left[
           			\ln
           			\left(
           			\dfrac{1+\sqrt{1-1/x}}
           			{1-\sqrt{1-1/x}}
           			\right)
           			-i\pi
           			\right]^2,
           			& x>1.
           		\end{cases}
           	\end{aligned}
           \end{equation}

	\bibliography{}

\end{document}